\documentclass[a4paper,12pt]{article}
\usepackage{amsmath}
\usepackage[pdftex]{graphicx}
\usepackage{amsmath}
\usepackage{amsmath,amsthm}
\usepackage{amsfonts}
\def\({\left(}
\def\){\right)}
\def\DS{\displaystyle}
\def\tr{{\rm tr}}
\def\av#1{\Bigl\langle{#1}\Bigr\rangle}
\def\be#1{\begin{equation}\label{#1}}
\def\ee{\end{equation}}
\def\BM{\begin{bmatrix}} % brackets [...]
\def\EM{\end{bmatrix}}

\def\[{\left[}
\def\]{\right]}

\def\av#1{\Bigl\langle{#1}\Bigr\rangle}
\def\dfeq{\,\stackrel{\mbox{\scriptsize def}}{=}\,}

\newcommand{\Vect}[1]{\boldsymbol{\mathrm{#1}}{}}
\newcommand{\Matr}[1]{\boldsymbol{{#1}}{}}

\def\eq#1{(\ref{#1})}

\def\xx{\Vect{x}}

\def\TT{\Matr{T}}
\def\DD{\Matr{D}}

\def\pp{\Vect{p}}

\def\kk{\Vect{k}}
\def\yy{\Vect{y}}
\def\vv{\Matr{v}}

\def\Omb{\boldsymbol{\varOmega}} %\boldsymbol\varOmega
\def\Lamb{\boldsymbol{\varLambda}}

\def\va{\Vect{a}_{\alpha}}

\def\paptit#1{#1}

\def\Xi{\Matr{\xi}{}}

\def\MM{\Matr{M}{}}

\def\Gamma{\Matr{\gamma}{}}
\def\PP{\Matr{P}{}}

\def\ii{{\rm i}}

\def\xx{\Vect{x}}

\def\EE{\Matr{E}}

\def\u{\Matr{u}}

\def\CC{\Matr{C}}

\def\yy{\Vect{y}}
\def\vv{\Vect{v}}

\def\va{\Vect{a}_{\alpha}}

\def\be#1{\begin{equation}\label{#1}}
\def\ee{\end{equation}}
\def\eq#1{(\ref{#1})}

\def\oa{\end{array} $$}

\let\DS=\displaystyle

\def\({\left(}
\def\){\right)}

\let\DS     = \displaystyle

\def\v {\Vect{v}}

\def\vv{\Vect{v}}

\def\va{\Vect{a}_{\alpha}}

\def\xx{\Vect{x}}

\def\EE{\Matr{E}}

\def\yy{\Vect{y}}
\def\vv{\Matr{v}}

\def\ve{\Vect{e}}

\def\va{\Vect{a}_{\alpha}}

\def\<{\langle}
\def\>{\rangle}

\def\Xi{\Matr{\xi}{}}

\begin{document}
\baselineskip 16pt
\title{Unsteady ballistic heat transport in infinite harmonic crystals}
\author{Vitaly A. Kuzkin\footnote{Peter the Great Saint Petersburg Polytechnical University;
              Institute for Problems in Mechanical Engineering RAS;
              e-mail: kuzkinva@gmail.com}}
\maketitle
%\rightline{\today}
%\bigskip
%\centerline{
%\vspace{10mm}

%\def\sect#1{\section{#1}}
%\def\sect#1{\paragraph{#1.}}

\begin{abstract}
We study thermal processes in infinite harmonic crystals having a unit cell with arbitrary number of particles. Initially particles have zero displacements and random velocities, corresponding to some initial temperature profile. Our main goal is to calculate spatial distribution of kinetic temperatures, corresponding to degrees of freedom of the unit cell, at any moment in time.  An approximate expression for the temperatures is derived from solution of lattice dynamics equations. It is shown that the temperatures are represented as a sum of two terms. The first term describes high-frequency oscillations of the temperatures caused by local transition to thermal equilibrium at short times. The second term describes slow changes of the temperature profile caused by ballistic heat transport. It is shown, in particular, that local values of temperatures, corresponding to degrees of freedom of the unit cell, are generally different even if their initial values are equal.  Analytical findings are supported by  results of numerical solution of lattice dynamics equations for diatomic chain and graphene lattice. Presented theory may serve for description of unsteady ballistic heat transport in real crystals with low concentration of defects. In particular, solution of the problem with sinusoidal temperature profile can be used for proper interpretation of experimental data obtained by the transient thermal grating technique.

\end{abstract}

\newpage
\tableofcontents
\newpage
\section{Introduction}
At macro scale, heat transport in solids is usually diffusive and well-described by the Fourier law. The law states that heat flux is proportional to temperature gradient with proportionality coefficient refereed to as the thermal conductivity.
Recent experiments for materials with low defect concentration have show that at micro and nano scale heat propagates ballistically~\cite{Cahill 2003 review, Chang review exp,  Hsiao ballistic, Pumarol ballistic}. In particular, it is demonstrated for many materials including nanowires~\cite{Anufriev nanowires, Hsiao ballistic}, nanotubes~\cite{Chang nanotubes}, graphene~\cite{Balandin graphene, Nika Balandin graphene, Xu grap exp 2014}, silicon membranes~\cite{Johnson Temp Grat} etc. that thermal conductivity strongly depend on sample size and the Fourier law is violated. Therefore development of theoretical models describing  ballistic heat transport is required.

Several approaches for derivation of equations describing ballistic heat transport are available in literature. In continuum theories, the equations are usually formulated using phenomenological approach. One of the phenomenological equations, describing wave nature of heat transport is the hyperbolic heat transfer equation, also refereed to as the Maxwell-Cattaneo-Vernotte equation~\cite{Chandrasekharaiah hyperbolic, Tzou hyperbolic}. Development of  phenomenological models is limited by small amount of available experimental data on unsteady ballistic heat transport. Another approach is based on Boltzmann transport equation~(BTE), formulated for distribution function of phonons~\cite{Kubo 1973, Pierls 1929}. Given known the distribution function, the temperature field can be calculated. The BTE is usually simplified using the relaxation time approximation~\cite{BGK, Klemens 1951, Kubo 1973} for the collision term. It allows to solve the BTE numerically~\cite{Hua Minnich 2014 BTE, Minnich BTE, Romano Grossmann} and to derive heat conduction equations~\cite{Gang Chen, Koh BTE, Mahan nonlocal BTE, Xu Hu 2010 balst-diff BTE}. In both cases, additional approximations are introduced~\cite{Sinha Goodson 2005 review BTE}. In particular, contribution of optical vibrations to heat transport is often neglected. Comprehensive review on application of the BTE to simulation of thermal transport can be found, e.g. in review papers~\cite{Cahill 2003 review, Lepri 2003, Sinha Goodson 2005 review BTE}. In the present paper, we use another approach for description of ballistic heat transport. The expression for evolution of temperature field is derived directly from lattice dynamics equations in harmonic approximation. This approach allows to describe heat transport in ballistic limit taking into account all important features of discrete system.

Analysis of heat transport in discrete systems is usually carried out in the so-called nonequilibrium steady-state. In this case, a material is kept between two thermostats with different temperatures. Given known the difference of temperatures, distance between thermostats, and a heat flux, one can calculate the effective heat conductivity of a material. This statement of the problem is widely used in both analytical studies~\cite{Bonetto2004, Lebowitz1967, Politi 2008} and computer simulations~\cite{DharSaito2016, Dhar altern mass, Lepri 2003, Xiong2013 diff stiff} of heat transport. Comprehensive reviews of results obtained in steady-state formulation are given e.g. in papers~\cite{Bonetto 2000, Dhar 2008, Lepri 2003} Calculating heat conductivity as a function of a sample length, one can determine conditions corresponding to ballistic and diffusive heat transport regimes. However, the steady-state formulation does not address the issue of temperature field evolution. Additionally, results of steady-state simulations significantly depend on the type of thermostat being used~\cite{Hoover thermostat, Dhar altern mass}. Therefore in the present paper we consider unsteady ballistic heat transport.

One of the goals of unsteady heat transport simulations is to describe time evolution of initial temperature profile. The initial profile field can be prescribed by assigning random initial velocities to particles. Then no thermostat is needed. The heat transport calculations are usually carried out numerically using, for example, molecular dynamics method~\cite{Gendelman 2010 nonstat, Krivtsov_LeZakharov, KosevichSavin, PiazzaLepri, Tsai nonstat}. The method allows to use realistic interatomic potentials and to consider effect of nonlinearity~(anharmonicity), defects, interfaces, and other features of real systems, which are hard to describe analytically. Though numerical modeling is a powerful tool for investigation of heat transport, many issues can not be addressed numerically. In particular, for crystals with several branches of dispersion relation, in numerical simulations it is hard to distinguish between contributions of acoustic and optical vibrations to the heat transport. Therefore analytical studies of unsteady heat transport are of great importance.

A promising model for investigation of ballistic heat transport is a harmonic crystal, i.e. a set of material points forming a perfect crystal lattice and interacting via linearized forces. In this model, harmonic waves do not interact with each other and therefore the heat transport is purely ballistic.
Unsteady ballistic heat transport in harmonic crystals is investigated  e.g. in papers~\cite{Babenkov2018, Gavrilov 2018 - 1, Guzev2018, Harris, Krivtsov DAN 2015, Kuzkin JPhys, Mielke, Sokolov}. In paper~\cite{Krivtsov DAN 2015}, an equation, refereed to as the ballistic heat equation,  describing evolution of temperature field in a one-dimensional chain with interactions of the nearest neighbors is derived. The equation is also valid~\cite{Babenkov2018} for one-dimensional chain with harmonic on-site potential~(elastic foundation). An expression for temperature field in scalar lattices with one degree of freedom per unit cell is derived in paper~\cite{Kuzkin JPhys}. Similar results are obtained by entirely different means in papers~\cite{Harris, Mielke}. In realistic systems, each unit cell usually has several degrees of freedom. To our knowledge, no closed-form expressions describing evolution of temperature field in crystals with several degrees of freedom per unit cell are available in literature. Therefore such expression is derived in the present paper.

The paper is organized as follows. In section~\ref{sect EM}, equations of motion for harmonic crystals are represented in a general form valid for one-, two-, and three-dimensional lattices with arbitrary number of particles per unit cell. Random initial conditions corresponding to an initial temperature profile are formulated. In section~\ref{sect exact sol} an exact solution of equations of motion is derived. The solution is used for calculation of temperatures, corresponding to different degrees of freedom. The temperatures are defined in section~\ref{sect Temp matr}.  In section~\ref{sect exact formula temp matr}, an exact expression describing temporal and spatial evolution of the temperatures is derived. In section~\ref{sect Approx formula}, simple approximate formula for the temperatures is derived. In section~\ref{sect special cases}, analysis of specific initial temperature profiles is presented. In particular, decay of a spatially sinusoidal profile of initial temperature is considered in section~\ref{sect sin general}. This problem is important, because it is closely related to experimental technique refereed to as the transient thermal grating~(TTG)~\cite{Johnson Temp Grat, Rogers Therm Grating, Huberman graphite experiment}. In the framework of TTG, a sinusoidal initial temperature profile is generated using the interference of two laser pulses. Decay of temperature profile yields information about  thermal properties of a material. Results obtained in sections~\ref{sect exact sol}-\ref{sect special cases} are applicable to crystals with an arbitrary lattice. In sections~\ref{sect Diatomic}, \ref{sect graphene} the general theory is employed for analysis of two particular cases, namely one-dimensional diatomic chain and graphene lattice. Analytical results are compared with numerical solution of lattice dynamics equations.

\section{Equations of motion of a crystal}\label{sect EM}
We consider infinite crystals in $d$-dimensional space, $d=1,2,3$. Unit cell of a crystal contains arbitrary number of particles.
In this section, we represent equations of motion of the unit cell in a matrix form~\cite{Kuzkin_2018_arxive}, convenient for analytical derivations.

Unit cells  of the lattice
are identified by their position vectors,~$\xx$\footnote{For analytical derivations, position vectors are more convenient than indices,
because number of indices depends on space dimensionality.}, in the undeformed state.
Each unit cell has~$N$ degrees of
freedom~$u_i(\xx), i=1,..,N$, corresponding to components of particle displacements~\footnote{$N$ is equal to number of particles in the unit cell multiplied by number of degrees of freedom per particle.}.
The displacements form a column:
\be{}
    \u(\xx) = \BM u_1 & u_2 & ... & u_N\EM^{\top},
\ee
where $\top$ stands for the transpose sign.

{\bf Remark.} Here and below matrices  are denoted by bold italic symbols, while invariant vectors, e.g. position vector, are denoted by bold symbols.

\begin{figure*}[htb]
\begin{center}
\includegraphics*[scale=0.36]{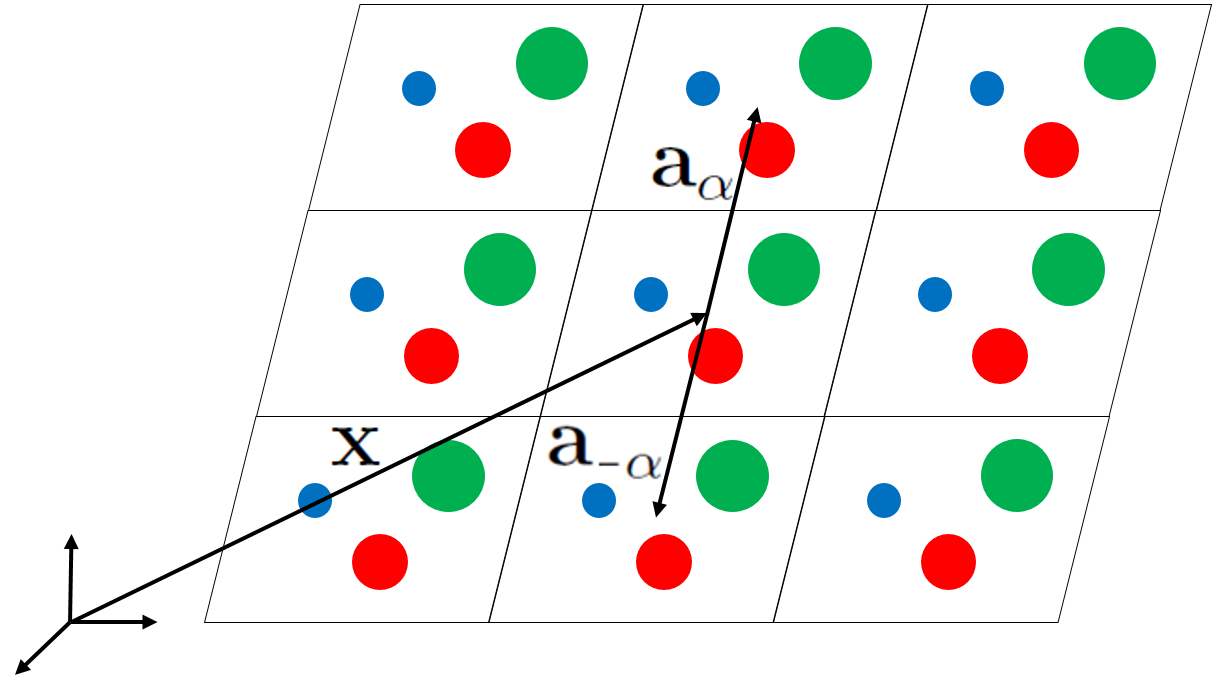}
\caption{Example of a complex two-dimensional lattice with three sublattices.
Particles forming sublattices have different color and size.}
\label{lattice}
\end{center}
\end{figure*}

Particles from the cell~$\xx$ interact with each other and with particles from neighboring unit cells, numbered by index $\alpha$. Vector connecting the cell~$\xx$ with neighboring cell number~$\alpha$ is denoted~$\va$.
Centers of unit cells always form a simple lattice\footnote{A lattice is referred to as simple lattice, if it coincides with itself under shift by a vector connecting any two particles.}, therefore numbering can be carried out so that vectors $\va$
satisfy the identity:
\be{}
 \va = -\Vect{a}_{-\alpha}.
\ee
Here~$\Vect{a}_{0} = 0$. Vectors $\va$ for a sample lattice are shown in figure~\ref{lattice}.

In the present paper, an infinite crystal is considered as a limiting case of a crystal under periodic boundary conditions. A periodic cell contains~$n^d$ unit cells~($n$ cells in each direction). 
Displacements of particles satisfy periodic boundary conditions:
\be{PBC}
    \u(\xx) = \u\(\xx + \sum_{j=1}^d C_j n \Vect{b}_j\),
\ee
where~$\Vect{b}_j$ are primitive vectors of the lattice; $C_j$ are integers. Further analytical derivations are carried out for~$n \rightarrow \infty$, while in computer simulations $n$ is finite.

We represent equations of motion in a quite general form, applicable to one-, two-, and three-dimensional lattices with arbitrary number of degrees of freedom per unit cell. In harmonic crystals, the total force acting on each particle is represented as a linear combination of displacements of
all other particles. Using this fact, we write equations of motion in the form~\cite{Kuzkin_2018_arxive, Mielke}:
\be{EM matr}
\begin{array}{l}
\DS \MM \dot{\vv}(\xx) = \sum_{\alpha} \CC_{\alpha} \u(\xx + \va), \qquad \CC_{\alpha} = \CC_{-\alpha}^{\top},
\end{array}
\ee
where~$\vv=\dot{\u}$; $\u(\xx + \va)$ is a column of displacements of particles from unit cell~$\alpha$;  $\MM$ is diagonal $N \times N$ matrix composed of particles' masses; coefficients of matrix~$\CC_{\alpha}$ determine stiffnesses of springs connecting particles from unit cell~$\xx$ with particles from neighboring cell~$\alpha$; matrix~$\CC_{0}$ describes interactions of particles inside the unit cell~$\xx$\footnote{Additionally, matrix~$\CC_0$ can include stiffnesses of harmonic on-site potential.}. Summation is carried out with respect to all unit cells~$\alpha$, interacting with unit cell~$\xx$~(including~$\alpha=0$).

{\bf Remark.} Relation~$\CC_{\alpha} = \CC_{-\alpha}^{\top}$ guarantees that dynamical matrix of the lattice~\eq{Omega} is Hermitian~(see section~\ref{sect exact sol} for more details).

Formula~\eq{EM matr} describes motion of monoatomic and polyatomic crystals in one-, two-~, and three-dimensional cases.  For~$N=1$~(one degree of freedom per unit cell), equation~\eq{EM matr}
governs dynamics of the so-called scalar lattices\footnote{In scalar lattices each particle has only one degree of freedom. This model is applicable to monoatomic one-dimensional chains with interactions of arbitrary number of neighbors and to out-of-plane motions of monoatomic two-dimensional lattices.}, considered, for example, in papers~\cite{Harris, Kuzkin JPhys, Mielke}. Monoatomic two- and three-dimensional lattices are covered if we put~$N=d$ and $\CC_{\alpha}=\CC_{\alpha}^{\top}$ in formula~\eq{EM matr}. In the present paper, for illustration we consider polyatomic lattices with two particles per unit cell, namely one-dimensional diatomic  chain~(section~\ref{sect Diatomic}) and out-of-plane motions of graphene lattice~(section~\ref{sect graphene}).

\section{Initial conditions}
In this section, we specify initial conditions for particles, corresponding to initial temperature profile.

The following initial conditions, typical for molecular dynamics
modeling~\cite{Allenbook}, are used:
\be{IC matrix0}
 \u(\xx) = 0, \quad \vv(\xx) = \vv_0(\xx),
\ee
Here  $\vv_0(\xx)$ is a column of random initial velocities of particles from unit cell~$\xx$ such that
\be{IC covar}
    \av{\vv_0(\xx)} = 0, \qquad \av{\vv_0(\xx)\vv_0(\yy)} = \Matr{B}(\xx)\delta_D(\xx-\yy),
\ee
where $\av{...}$ stands for mathematical expectation\footnote{In computer simulations, mathematical expectation is approximated by average over realizations with different random initial conditions.};
 $\delta_D(0) = 1$; $\delta_D(\xx-\yy)=0$ for~$\xx\neq\yy$. In other words, components of~$\vv_0(\xx)$ are random numbers with zero mean\footnote{In this case mathematical expectations of all velocities are equal to zero at any moment in time.} and generally different variances given by diagonal elements of matrix~$\Matr{B}(\xx)$. Off-diagonal elements of matrix~$\Matr{B}(\xx)$ are equal to covariances of initial velocities, corresponding to different degrees of freedom of unit cell, $\xx$. Initial velocities of particles from different unit cells are statistically independent. 

From macroscopic point of view, initial conditions~\eq{IC matrix0}, \eq{IC covar} specify some initial temperature profile~(see formulas~\eq{matr temp}, \eq{kin temp}) and zero initial heat fluxes\footnote{Heat flux in a lattice is proportional to covariance of force and velocity~(see e.g. paper~\cite{Krivtsov DAN 2015}). Since initial displacements are equal to zero then forces and fluxes vanish.}. Examples of initial conditions~\eq{IC matrix0}, \eq{IC covar} are given by formulas~\eq{IC chain gen}, \eq{IC graph gen}. Initial conditions~\eq{IC matrix0}, \eq{IC covar} can be considered as a result of heating of a crystal by an ultrashort laser pulse~\cite{Huberman graphite experiment, Indeitsev2017, Johnson Temp Grat, Poletkin2012, Rogers Therm Grating}.

In the following section, an exact solution of equations of motion~\eq{EM matr} with initial conditions~\eq{IC matrix0} is obtained. The solution is employed for description of thermal processes, such as the ballistic heat transport.

\section{Exact solution of equations of motion}\label{sect exact sol}
In this section, we derive an exact solution of equation~\eq{EM matr}  with initial conditions~\eq{IC matrix0}, \eq{IC covar} using the discrete
Fourier transform with respect to components of position vector~$\xx$.\footnote{In fact, while using the discrete Fourier transform, we assume periodic boundary conditions in all directions and consider the limit of infinite size of the periodic cell.}

Position vector,~$\xx$, of a unit cell is represented as
\be{}
   \xx = \sum_{j=1}^d  z_j \Vect{b}_j,
\ee
where $\Vect{b}_j, j=1,..,d$ are primitive vectors of the lattice; $z_1,..,z_d$ are integer indices of the unit cell; $d$ is space dimensionality. Direct and inverse discrete Fourier transforms with respect to variables~$z_1, .., z_d$ for
an infinite lattice are defined as
\be{DFT}
\begin{array}{l}
\DS \hat{\u}(\kk)
  = \sum_{\xx} \u(\xx) e^{-\ii \kk \cdot \xx},
 \qquad
   \kk = \sum_{j=1}^d p_j  \tilde{\Vect{b}}_j,
 \\[4mm]
\DS \u(\xx) = \int_{\kk} \hat{\u}(\kk) e^{\ii \kk \cdot \xx} {\rm d} p_1.. {\rm d} p_d.
  \end{array}
\ee
Here $\hat{\u}(\kk)$ is Fourier image of $\u$; $\ii^2=-1$; $\kk$ is wave vector;  $\tilde{\Vect{b}}_j$ are vectors of the reciprocal
basis, i.e.~$\Vect{b}_i\cdot\tilde{\Vect{b}}_j = \delta_{ij}$, where $\delta_{ij}$ is the Kroneker delta.
For brevity,  the following notation is used:
\be{int k}
\int_{\kk} ... {\rm d}\kk
\dfeq
 \frac{1}{(2\pi)^d}\int_{0}^{2\pi}..\int_{0}^{2\pi} ... {\rm d} p_1.. {\rm d} p_d,
 \qquad
 \sum_{\xx}...
 \dfeq
 \sum_{z_1=-\infty}^{+\infty}...\sum_{z_d=-\infty}^{+\infty}...
\ee

Applying the discrete Fourier transform~\eq{DFT} to formulas~\eq{EM matr}, \eq{IC matrix0}, \eq{IC covar},
yields equation
\be{Omega}
 \MM^{\frac{1}{2}} \ddot{\hat{\u}} = -\Omb \MM^{\frac{1}{2}}\hat{\u},  \qquad
 \Omb(\kk) =- \sum_{\alpha} \MM^{-\frac{1}{2}}\CC_{\alpha}\MM^{-\frac{1}{2}} e^{\ii \kk\cdot\va},
\ee
with initial conditions
\be{IC u im}
  \hat{\u} = 0, \qquad \dot{\hat{\u}} = \hat{\vv}_0 = \sum_{\xx}\vv_0(\xx) e^{-\ii \kk \cdot \xx}.
\ee
Matrix~$\Omb$ in formula~\eq{Omega} is the {\it dynamical matrix} of the lattice~\cite{Dove}.

To simplify equation~\eq{Omega}, we use the fact that dynamical matrix~$\Omb$ is Hermitian, i.e. it is equal to its own conjugate transpose. Then it can be represented as
\be{PP def}
 \Omb = \PP\Lamb \PP^{*{\top}}, \qquad \Lambda_{ij} = \omega_j^2 \delta_{ij},
\ee
where  $\omega_j^2, j=1,..,N$ are eigenvalues of matrix~$\Omb$ and $\omega_j(\kk)$ are branches of dispersion relation for the lattice~(below we consider only nonnegative frequencies, i.e.~$\omega_j(\kk) \geq 0$); $*$ stands for complex conjugate; matrix $\PP$ is composed of normalized eigenvectors of matrix~$\Omb$\footnote{Matrix $\PP$ is unitary, i.e. $\PP \PP^{*{\top}} = \EE$,  where $\EE$ is unit matrix}.  The eigenvectors are referred to as the polarization vectors~\cite{Dove}.

{\bf Remark.} We assume that branches of dispersion relation do not intersect with each other, i.e. all eigenvalues of the dynamic matrix~$\Omb$ are distinct. The case of intersecting branches should be considered separately.

We substitute formula~\eq{PP def} into equation~\eq{Omega},
 multiply both parts by~$\PP^{*{\top}}$ and introduce new
 variable~$\Matr{w} = \PP^{*{\top}} \MM^{\frac{1}{2}}\hat{\u}$. Then we obtain a system of decoupled equations for elements~$w_j$ of
 vector~$\Matr{w}$:
\be{Omega PP}
    \ddot{\Matr{w}} = -\Lamb \Matr{w}
    \quad
    \Leftrightarrow
    \quad
    \ddot{w}_j = -\omega_j^2 w_j.
\ee
Solving these equations with initial conditions~\eq{IC u im} we obtain the following expression for~$\dot{\Matr{w}}$:
\be{}
 \dot{w}_j = \{\PP^{*\top} \MM^{\frac{1}{2}} \hat{\vv}_0\}_j \cos\(\omega_j t\) \quad
    \Leftrightarrow
    \quad
    \dot{\Matr{w}} = \DD \PP^{*\top} \MM^{\frac{1}{2}} \hat{\vv}_0, \qquad D_{ij}(\kk, t) = \cos\(\omega_j(\kk) t\)\delta_{ij}.
\ee
Here~$\{...\}_j$ stands for $j$th element of a column.
Then using definition of~$\Matr{w}$, we represent Fourier-images of velocities in the form
\be{hat v}
    \hat{\vv} = \MM^{-\frac{1}{2}} \PP \DD \PP^{*\top} \MM^{\frac{1}{2}} \hat{\vv}_0.
\ee
Applying the inverse discrete Fourier transform to formula~\eq{hat v},
yields the following expression for particle velocities:
\be{exact dyn}
 \vv(\xx) = \MM^{-\frac{1}{2}}\int_{\kk} \PP \DD \PP^{*{\top}} \MM^{\frac{1}{2}} \hat{\vv}_0 e^{\ii \kk\cdot\xx} {\rm d} \kk,
\ee
where~$\hat{\vv}_0$ is defined by formula~\eq{IC u im}.

Thus formula~\eq{exact dyn} is an exact solution of equation~\eq{EM matr}  with initial conditions~\eq{IC matrix0}. In the following sections, spatial distribution of temperature is calculated using solution~\eq{exact dyn}.

\section{Kinetic temperature. Temperature matrix}\label{sect Temp matr}
Since initial conditions~\eq{IC matrix0}, \eq{IC covar} are random then particle velocities given by formula~\eq{exact dyn} are also random. Following~\cite{Krivtsov 2014 DAN, Krivtsov DAN 2015, Krivtsov 2019 ballistic} we
consider an infinite number of realizations of initial conditions~\eq{IC matrix0}, \eq{IC covar}. It allows to introduce statistical characteristics such as kinetic temperature.

In harmonic crystals, kinetic temperatures, corresponding to degrees of freedom of the unit cell, are generally different~(see e.g. figures~\ref{Heaviside_chain_m1m2}, \ref{A11 chain}, \ref{A22 chain}).
Therefore in order to characterize thermal state of the unit cell, $\xx$, we
use~$N\times N$ matrix, $\TT(\xx)$, further referred
to as the {\it temperature matrix}~\cite{Kuzkin_2018_arxive}:
\be{matr temp}
 k_B\TT(\xx) = \MM^{\frac{1}{2}}\av{\vv(\xx) \vv(\xx)^{\top}}\MM^{\frac{1}{2}}
 \quad
 \Leftrightarrow
 \quad
  k_B T_{ij} = \sqrt{M_iM_j} \av{v_i v_j},
\ee
where~$\MM^{\frac{1}{2}}\MM^{\frac{1}{2}} = \MM$; $M_i$ is $i$-th element of matrix~$\MM$, equal to a mass corresponding to $i$-th degree of freedom of the unit cell;
$k_B$ is the Boltzmann constant; brackets~$\av{..}$ stand for mathematical expectation\footnote{In computer simulations, mathematical expectation is approximated by average over realizations with different random initial conditions.}.
Diagonal element, $T_{ii}$, of the temperature matrix is refereed to as kinetic temperature, corresponding to~$i$-th degrees of freedom of the unit cell. Off-diagonal elements, $T_{ij}$,  characterize correlation between  components,~$i, j$, of velocity column,~$\vv(\xx)$.

We also use conventional kinetic temperature, $T$, proportional
to the total kinetic energy of the unit cell:
\be{kin temp}
  T(\xx) = \frac{1}{N}\,\tr \TT(\xx),
\ee
where~$N$ is a number of degrees of freedom per unit cell, $\tr(..)$ stands
for the trace~(sum of diagonal elements). If initial kinetic energy is uniformly distributed among degrees of freedom of the unit cell then  kinetic temperatures, corresponding to all degrees of freedom of the unit cell, are equal to~$T$.

\section{Exact formula for the temperature matrix}\label{sect exact formula temp matr}
In this section, we derive an exact expression for the temperature matrix using the solution~\eq{exact dyn} of lattice dynamics equation.

To calculate temperature matrix, we substitute solution~\eq{exact dyn}
into definition~\eq{matr temp}~\footnote{Here the following identities were used: $\int_{\kk} \Matr{F}(\kk){\rm d} \kk \int_{\kk}
\Matr{F}^{*\top}(\kk){\rm d} \kk
= \int_{\kk_1}\int_{\kk_2} \Matr{F}(\kk_1)\Matr{F}^{*\top}(\kk_2)  {\rm d} \kk_1 {\rm d} \kk_2$ and $\v = \v^{*}$.}:
\be{TT 1}
k_B\TT(\xx) = \int_{\kk_1}\int_{\kk_2}
\PP_1 \DD_1 \PP_1^{*{\top}} \MM^{\frac{1}{2}}
\av{\hat{\vv}_0(\kk_1) \hat{\vv}_0(\kk_2)^{*{\top}}}
\MM^{\frac{1}{2}} \PP_2\DD_2\PP_2^{*\top}
e^{\ii \(\kk_1-\kk_2\)\cdot\xx}  {\rm d} \kk_1{\rm d} \kk_2.
\ee
Here and below~$\PP_j = \PP(\kk_j)$, $\DD_j  = \DD(\kk_j)$, $j=1,2$.

Initial conditions~\eq{IC matrix0}, \eq{IC covar} are such that initial velocities of any pair of unit cells~$\yy_1, \yy_2$ are uncorrelated.
Therefore the following identity  is satisfied~
\be{uncor vel}
 \av{\vv_0(\yy_1) \vv_0(\yy_2)^{*{\top}}} = \MM^{-\frac{1}{2}} k_B \TT_0(\yy_1) \delta_D(\yy_1-\yy_2)\MM^{-\frac{1}{2}},
\ee
Here~$\TT_0$ is the initial temperature matrix, which is 
calculated by substituting initial velocities,~$\vv_0(\xx)$, into formula~\eq{matr temp};
$\delta_D$ is defined after formula~\eq{IC covar}. Using identity~\eq{uncor vel}, we make the following transformations:
\be{cov in}
\begin{array}{l}
 \DS \av{\hat{\vv}_0(\kk_1) \hat{\vv}_0(\kk_2)^{*{\top}}}
 =
 \sum_{\yy_1, \yy_2} \av{\vv_0(\yy_1) \vv_0(\yy_2)^{*{\top}}}
 e^{-\ii \(\kk_1\cdot\yy_1 - \kk_2\cdot\yy_2\)}
 =
 \\[4mm]
 \DS
 %=  k_B \sum_{\yy_1, \yy_2} \MM^{-\frac{1}{2}} \TT_0(\yy_1) \delta(\yy_1-\yy_2)\MM^{-\frac{1}{2}}
 %e^{-\ii \(\kk_1\cdot\yy_1 - \kk_2\cdot\yy_2\)}
 = k_B \sum_{\yy} \MM^{-\frac{1}{2}} \TT_0(\yy) \MM^{-\frac{1}{2}}
 e^{-\ii \(\kk_1-\kk_2\)\cdot\yy}.
 \end{array}
\ee
Then substitution of formula~\eq{cov in} into~\eq{TT 1}, yields
\be{TT exact}
\TT(\xx, t) = \int_{\kk_1}\int_{\kk_2}\sum_{\yy}
\PP_1 \DD_1 \PP_1^{*{\top}} \TT_0(\yy)
 \PP_2\DD_2\PP_2^{*\top} e^{\ii \(\kk_1-\kk_2\)\cdot(\xx-\yy)}  {\rm d} \kk_1{\rm d} \kk_2.
\ee
Formula~\eq{TT exact} is {\it an exact} expression for the temperature matrix.

{\bf Remark.} Formula~\eq{TT exact} for the temperature matrix is symmetric with respect to time, i.e. invariant with respect to substitution~$t$ by $-t$. This fact follows from the same property of equations of motion~\eq{EM matr}.

The temperature matrix can be exactly represented as a sum of ``fast'' and ``slow'' terms:
\be{fast slow ex}
\begin{array}{l}
 \DS
\TT = \TT_F + \TT_S,
%\\[4mm]
%\DS
\qquad
\TT_F(\xx) = \int_{\kk_1}\int_{\kk_2}\sum_{\yy}
\PP_1 \TT'_F\PP_2^{*\top} e^{\ii \(\kk_1-\kk_2\)\cdot(\xx-\yy)}  {\rm d} \kk_1{\rm d} \kk_2,
\\[4mm]
\DS
\TT_S(\xx) = \int_{\kk_1}\int_{\kk_2}\sum_{\yy}
\PP_1 \TT'_S\PP_2^{*\top} e^{\ii \(\kk_1-\kk_2\)\cdot(\xx-\yy)}  {\rm d} \kk_1{\rm d} \kk_2, \\[4mm]
 \DS
 \{\TT'_F\}_{ij} =  \frac{1}{2}\{\PP_1^{*\top} \TT_0(\yy) \PP_2\}_{ij} \[\cos((\omega_i(\kk_1) + \omega_j(\kk_2))t)  + (1-\delta_{ij})\cos((\omega_i(\kk_1) - \omega_j(\kk_2))t)\],
\\[4mm]
\DS
 \{\TT'_S\}_{ij} =  \frac{1}{2}\{\PP_1^{*\top} \TT_0(\yy) \PP_2\}_{ij} \delta_{ij}\cos\((\omega_j(\kk_1) - \omega_j(\kk_2))t\).
\end{array}
\ee
Here~$\{...\}_{ij}$ is element~$i,j$ of the matrix.
The representation~\eq{fast slow ex} is based on the observation that~$\TT_F$ and $\TT_S$ have different characteristic time scales. The difference of time scales is most clearly demonstrated by example, considered in section~\ref{sect sin general}. Physical meaning of $\TT_F$ and $\TT_S$ is discussed in the next section.

It is seen that exact formulas~\eq{TT exact}, \eq{fast slow ex} for the temperature matrix are quite complicated and require intensive calculations~(double integration with respect to wave vectors~$\kk_1, \kk_2$ and summation with respect to all unit cells). Therefore in the following section we present a simple approximate formula for the temperature matrix.

\section{Approximate formula for the temperature matrix. Fast and slow thermal processes} \label{sect Approx formula}
The main result of the present paper is the following approximate formula
for the temperature matrix~(derivation is presented in the Appendix):
\be{fast and slow app}
\begin{array}{l}
\DS
\TT = \TT_F + \TT_S,
\qquad
\TT_F \approx \int_{\kk}\PP \tilde{\TT}_F\PP^{*\top} {\rm d} \kk,
\qquad
\DS \TT_S \approx \int_{\kk}
\PP \tilde{\TT}_S \PP^{*\top}  {\rm d} \kk,
\\[4mm]
 \DS
\{\tilde{\TT}_F\}_{ij} = \frac{1}{2}\{\PP^{*\top} \TT_0(\xx) \PP\}_{ij} \[\cos((\omega_i + \omega_j)t)  + (1-\delta_{ij})\cos((\omega_i - \omega_j)t)\],
 \\[4mm]
 \DS \{\tilde{\TT}_S\}_{ij} = \frac{1}{4}\{\PP^{*\top} \(\TT_0(\xx + \Vect{v}_g^j t) +  \TT_0(\xx - \Vect{v}_g^j t)\)\PP\}_{jj}\delta_{ij}.
\end{array}
\ee
Here $\PP=\PP(\kk)$; $\Vect{v}_g^j$ is the group velocity corresponding
to~$j$-th branch of dispersion relation,~$\omega_j$\footnote{For each value of the wave vector,~$\kk$,
and for each branch,~$j$, there are positive and negative frequencies~(except~$\omega_j=0$).
In formula~\eq{fast and slow app}, positive frequencies~$\omega_j, j=1,..,N$ are used.}:
\be{def group vel}
  \Vect{v}_{g}^{j} = \frac{{\rm d} \omega_{j}}{{\rm d} \kk}
  = \sum_{i=1}^d  \frac{\partial \omega_{j}}{\partial p_i} \Vect{b}_i, \qquad \kk = \sum_{i=1}^d p_i  \tilde{\Vect{b}}_i.
\ee
Since formula~\eq{fast and slow app} contains only a single integral with respect to~$\kk$, it is  significantly more convenient
for analysis and calculations than exact expression~\eq{fast slow ex}.

{\bf Remark.} Function~$\TT_0$ is originally defined on a discrete set of position vectors, $\xx$, of unit cells~(see formula~\eq{matr temp}), while argument of this function~$\xx \pm \Vect{v}_g^j t$ changes in space continuously. Therefore further we assume that~$\TT_0$ can be defined for the whole space in such a way that at points~$\xx$ is coincides with values given by formula~\eq{matr temp}.

The first term, $\TT_F$, in formula~\eq{fast and slow app} describes short time behavior of the temperature matrix~(fast process~\cite{Krivtsov 2014 DAN, Kuzkin JPhys}). At short times, the temperature matrix oscillates.  The oscillations are caused by redistribution of energy among kinetic and potential forms and redistribution of energy among degrees of freedom of the unit cell. These oscillations at different spatial points are independent. At large time scale~$\TT_F$ tends to zero.

The second term, $\TT_S$, in formula~\eq{fast and slow app} describes large time behavior of the temperature matrix~(slow process~\cite{Krivtsov DAN 2015, Krivtsov 2019 ballistic, Kuzkin JPhys}). At large time scale, changes of the temperature matrix are caused by ballistic heat transport. The temperature matrix is represented as a superposition of waves traveling with group velocities~$\Vect{v}_g^j(\kk)$. Shapes of the waves are determined by initial temperature profile~$\TT_0$.

{\bf Remark.} Approximate formula~\eq{fast and slow app} for the
temperature matrix have the same property as the exact formula~\eq{TT exact}. It is symmetric with respect to time, i.e. substitution~$t$ by $-t$. At the same time, analysis of formula~\eq{fast and slow app} and results of numerical simulations suggest that thermal processes in infinite harmonic crystals are irreversible.

{\bf Remark.} According to formula~\eq{fast slow ex}, the temperature matrix at the unit cell~$\xx$ depends
on initial temperatures at all other unit cells. This fact is a consequence of infinite propagation speed of disturbances in discrete systems described by equations of motion~\eq{EM matr}. In contrast, approximate formula~\eq{fast and slow app} does not have this artifact. According to  formula~\eq{fast and slow app}, temperature matrix of the unit cell~$\xx$ at time~$t$
depends on initial temperature matrices of unit cells which are
not farther from cell~$\xx$ than~$\max_{\kk, j}\(|\Vect{v}_g^j|t\)$.

{\bf Remark.} Comparison of formula~\eq{fast and slow app} with results of paper~\cite{Kuzkin_2018_arxive} shows that $\TT_S(\xx)$ at~$t=0$ is equal to the equilibrium value of the temperature matrix in the uniformly heated crystal with initial temperature matrix~$\TT_0(\xx)$.

\section{Specific profiles of initial temperature}\label{sect special cases}

In this section, we apply general solution~\eq{fast and slow app} to several particular  initial temperature profiles. The results are valid for all lattices described by equations of motion~\eq{EM matr}.

\subsection{Uniform initial temperature profile}
In this subsection, we consider spatially uniform distribution of initial temperature~($\TT_0$ is independent of~$\xx$). In this case changes of the temperature matrix are caused by two physical processes: redistribution of energy between kinetic and potential forms and redistribution of energy between degrees of freedom. These processes are usually observed in molecular dynamics simulations~\cite{Allenbook}. Analytical description of these processes for several specific one- and two-dimensional lattices is presented in papers~\cite{Babenkov2016, Krivtsov 2014 DAN, Kuzkin JPhys, Kuzkin_DAN, Kuzkin_FTT}. Generalization for the case of polyatomic crystals is carried out in paper~\cite{Kuzkin_2018_arxive}. In these works, the approach, originally proposed in paper~\cite{Krivtsov 2014 DAN} and based on analysis of velocity covariances, is used. Here we show that identical results follow from formula~\eq{fast and slow app} derived using solution of lattice dynamics equations.

In the case of spatially uniform distribution of temperature matrix,
formula~\eq{fast and slow app} reads
\be{T unif sol}
\begin{array}{l}
\DS
    \TT_F = \int_{\kk}\PP \tilde{\TT}_F\PP^{*\top} {\rm d} \kk,
    \qquad
    \TT_S = \int_{\kk}\PP \tilde{\TT}_S \PP^{*\top}  {\rm d} \kk,
\\[4mm]
 \DS
\{\tilde{\TT}_F\}_{ij} = \frac{1}{2}\{\PP^{*\top} \TT_0 \PP\}_{ij} \[\cos((\omega_i + \omega_j)t)  + (1-\delta_{ij})\cos((\omega_i - \omega_j)t)\].
\\[4mm]
\DS
  \{\tilde{\TT}_S\}_{ij} = \frac{1}{2}\{\PP^{*\top}\TT_0\PP\}_{jj}\delta_{ij}.
\end{array}
\ee
Expressions~\eq{T unif sol} coincide with exact formulas for the temperature matrix, obtained in paper~\cite{Kuzkin_2018_arxive} by entirely different means. In particular, the expression for~$\TT_S$ coincides with equilibrium  value of the temperature matrix~\cite{Kuzkin_2018_arxive}.

Therefore in the case of spatially uniform distribution of temperature, formula~\eq{fast and slow app} is {\it exact}.

\subsection{Initial equipartition}
In this section, we consider the case, when initial kinetic temperatures, corresponding to all degrees of freedom of the unit cell, are equal. Then initial temperature matrix is isotropic\footnote{Matrix is called isotropic if it is diagonal and all elements on the diagonal are equal.}, i.e.~$\TT_0 = T_0(\xx) \EE$, where~$\EE$ is the unit matrix.
Substitution of this expression into formula~\eq{fast and slow app} yields:
\be{TT gen spher}
\begin{array}{l}
 \DS \TT_F = \int_{\kk}
\PP \tilde{\TT}_F \PP^{*\top}  {\rm d} \kk,
 \qquad
 \TT_S = \int_{\kk}
\PP \tilde{\TT}_S \PP^{*\top}  {\rm d} \kk,
   \\[4mm]
 \DS
 \{\tilde{\TT}_F\}_{ij}  = \frac{1}{2}T_0(\xx)\delta_{ij} \cos(2\omega_j t),
 \qquad
 \{\tilde{\TT}_S\}_{ij}  = \frac{1}{4}\(T_0(\xx + \Vect{v}_g^j t) +  T_0(\xx - \Vect{v}_g^j t)\)\delta_{ij}.
  \end{array}
\ee
The kinetic temperature, proportional to trace of the temperature matrix~\eq{TT gen spher}, has form
\be{T tr}
\begin{array}{l}
 \DS T = T_F + T_S,
 \qquad
 T_F = \frac{T_0(\xx)}{2N} \sum_{j=1}^N\int_{\kk} \cos(2\omega_j t)
 {\rm d}\kk,
 \\[4mm]
 \DS T_S = \frac{1}{4N}\sum_{j=1}^N\int_{\kk}\(T_0(\xx + \Vect{v}_g^j t) +  T_0(\xx - \Vect{v}_g^j t)\) {\rm d} \kk.
  \end{array}
\ee

{\bf Remark.}  For scalar lattices~($N=1$), formula~\eq{T tr} coincides with the result obtained in paper~\cite{Kuzkin JPhys}. In paper~\cite{Kuzkin JPhys} the expression for temperature is derived by approximate solution of equation for covariances of velocities, while in the present paper it is derived from solution of lattice dynamics equations.

{\bf Remark.} Expression for~$T_S$ in  formula~\eq{T tr} is also consistent with results obtained in paper~\cite{Mielke} by entirely different means.
In paper~\cite{Mielke}, the expression for the total energy of the unit cell at large times is derived. At large times, kinetic and potential energies equilibrate~\cite{Kuzkin_2018_arxive} and
therefore behavior of the total energy and kinetic temperature are similar.

Formula~\eq{TT gen spher} shows that the temperature matrix for $t>0$ is generally {\it not isotropic}, i.e. temperatures, corresponding to degrees of freedom of the unit cell, are different even though initially they are equal. Further this important fact is illustrated by figures~\ref{Heaviside_chain_m1m2}, \ref{A11 chain}, \ref{A22 chain}.

\subsection{Fundamental solution}

In this subsection, we derive fundamental solution of ballistic heat transport problem.

The following spatial distribution of
initial temperature matrix is considered
\be{}
 \TT_0(\xx) = A\delta(\xx)\EE,
\ee
where $\delta(\xx)$ is Dirac delta function; $A$ is a constant.
Distribution of kinetic temperature at large times is given by formula~\eq{T tr}:
\be{T fund prel}
  T \approx T_S
  =
  \frac{A}{4N}\sum_{i=1}^N\int_{\kk}\(\delta(\xx + \Vect{v}_g^i t) + \delta(\xx - \Vect{v}_g^i t)\){\rm d} \kk.
\ee
Non-zero contribution to integral~\eq{T fund prel} comes from values,~$\kk_{ij}^*$,
of wave-vector such that argument of the delta function vanishes:
\be{eq for k}
    \Vect{v}_g^i(\kk_{ij}^*)  = \frac{\xx}{t}
    \qquad
        \mathrm{or}
    \qquad
    \Vect{v}_g^i(\kk_{ij}^*)  = -\frac{\xx}{t}.
\ee
Here~$j=1,..,n_i$, where $n_i$ is the number of real roots of equation~\eq{eq for k}, for $i$th branch or dispersion relation.
Then calculation of integrals in formula~\eq{T fund prel} yields
\be{T fund sol}
\begin{array}{l}
 \DS T = \frac{A}{4N(2\pi)^d t^d}\sum_{i=1}^N \sum_{j=1}^{n_i}\frac{1}{|\det \Matr{G}_i^d(\kk_{ij}^*)|},
 \end{array}
\ee
where summation is carried out with respect to all real roots~$\kk_{ij}^*, i=1,..,N, j=1,..,n_i$ of equations~\eq{eq for k}\footnote{Components of $\kk_{ij}^*$ belong to interval $[0; 2\pi]$.}; $\det$ stands for determinant of a matrix;
 $\Matr{G}_i^d$ is the Jacobian matrix in $d$-dimensional case:
\be{}
\Matr{G}_i^{1} = \frac{\partial v_{g}^j}{\partial p},
\qquad
    \Matr{G}_i^{2} =
  \BM
    \frac{\partial v_{gx}^i}{\partial p_1}  &  \frac{\partial v_{gx}^i}{\partial p_2} \\
    \frac{\partial v_{gy}^i}{\partial p_1}  &  \frac{\partial v_{gy}^i}{\partial p_2}
    \EM,
    \qquad
  \Matr{G}_i^{3} =
  \BM
   \frac{\partial v_{gx}^i}{\partial p_1}   &  \frac{\partial v_{gx}^i}{\partial p_2}  & \frac{\partial v_{gx}^i}{\partial p_3}\\
    \frac{\partial v_{gy}^i}{\partial p_1}   &  \frac{\partial v_{gy}^i}{\partial p_2}  & \frac{\partial v_{gy}^i}{\partial p_3}\\
   \frac{\partial v_{gz}^i}{\partial p_1}   &  \frac{\partial v_{gz}^i}{\partial p_2}  & \frac{\partial v_{gz}^i}{\partial p_3}
    \EM,
\ee
where~$v_{gx}^i, v_{gy}^i, v_{gz}^i$ are components of vector of group velocity~$\Vect{v}_g^i(\kk)$.

Since the group velocity is usually bounded then from formulas~\eq{eq for k}, \eq{T fund sol}, it follows that the heat front, corresponding to
fundamental solution, is a $d$-dimensional sphere given by
\be{front eq}
\begin{array}{l}
  |\xx|  = t \max_{\kk, i}|\Vect{v}_{g}^i|.
\end{array}
\ee
In other words, the heat front propagates with maximum group velocity. This fact is illustrated in figure~\ref{graphene_delta_all}.

{\bf Remark.} For one-dimensional and two-dimensional scalar lattices~($N=1$) fundamental solutions are
obtained in papers~\cite{Krivtsov DAN 2015, Kuzkin JPhys}. These solutions coincide with particular cases of formula~\eq{T fund sol}.

{\bf Remark.} Fundamental solution~\eq{T fund sol} can be used for calculation of temperature field in a crystal subjected to point heat supply of constant intensity. In paper~\cite{Gavrilov 2018 - 1} it is shown that the temperature field is equal to integral of the fundamental solution with respect to time.

\subsection{Thermal contact of cold and hot half-spaces}

Consider thermal contact of two half-spaces with initial temperatures~$T_b$
and~$T_b + \Delta T$. This problem is important, because
it is closely related to classical definition of temperature~\cite{Hoover_stat_phys}.
By the definition, temperatures of two bodies in thermodynamic
equilibrium are equal. The problem considered below
demonstrates the transition to thermodynamic equilibrium.

The initial temperature distribution in direction,
given by unit vector~$\Vect{e}$, has form:
\be{ICs Heaviside}
  \TT_0 = \(T_b +  \Delta T H(x)\)\EE, \qquad x = \Vect{\xx} \cdot \Vect{e},
\ee
where~$H$ is the Heaviside function.
Substituting formula~\eq{ICs Heaviside} into~\eq{TT gen spher}, yields
\be{TfTs Heaviside}
\begin{array}{l}
 \DS \{\tilde{\TT}_F\}_{ij} = \frac{1}{2}\(T_b + \Delta T H(x)\)\delta_{ij} \cos(2\omega_j t),
 \\[4mm]
 \DS \{\tilde{\TT}_S\}_{ij} = \frac{T_b}{2}\delta_{ij} + \frac{\Delta T}{4}\(H\(x + \Vect{v}_g^j \cdot \Vect{e} t\) + H\(x - \Vect{v}_g^j \cdot \Vect{e} t\)\)\delta_{ij}.
 \end{array}
\ee
Computing kinetic temperature by formulas~\eq{TT gen spher}, \eq{TfTs Heaviside} and using
the property of Heaviside function~$H(ax)=H(x)$, yields
\be{TfTs Heaviside trace}
\begin{array}{l}
 \DS T = T_F + T_S,
 \qquad
 T_F = \frac{1}{2N}\(T_b + \Delta T H(x)\)\sum_{j=1}^N\int_{\kk} \cos(2\omega_j t){\rm d}\kk,
 \\[4mm]
 \DS T_S\(\frac{x}{t}\) = \frac{T_b}{2} + \frac{\Delta T}{4N}\sum_{j=1}^N\int_{\kk}\(H\(\frac{x}{t} + \Vect{v}_g^j \cdot \Vect{e}\) + H\(\frac{x}{t} - \Vect{v}_g^j \cdot \Vect{e}\)\){\rm d}\kk.
 \end{array}
\ee
Formula~\eq{TfTs Heaviside trace} shows that slow part of the temperature matrix,~$T_S$, is {\it self-similar}, i.e. dependent on~$x/t$. This fact is used for comparison with results of numerical solution of equations of motion in sections~\ref{sect thermal contact chain}, \ref{sect thermal contact graphene}.

\subsection{Sinusoidal initial temperature profile:
application to transient thermal grating}\label{sect sin general}
In this subsection, we consider spatially sinusoidal profile of initial temperature. This problem is closely related to transient thermal grating technique~\cite{Johnson Temp Grat, Rogers Therm Grating}. In the framework of this experimental technique, the sinusoidal profile in a thin film is generated using the interference of two laser pulses. Amplitude of the temperature profile decays in time due to heat transport. Measurement of the amplitude yields information on thermal properties of a material. Here we present an analytical solution of this problem, corresponding to purely ballistic regime of heat transport.

The initial temperature profile in direction given
by unit vector~$\Vect{e}$ has form:
\be{T0 sin}
 \TT_0(\xx) = \(T_b + \Delta T \sin\frac{2\pi x}{L}\) \EE, \qquad x = \xx \cdot \Vect{e},
\ee
where $T_b$, $\Delta T$ are constants; $\Delta T < T_b$; $L$ is length of the periodic cell. Note that initial temperatures, corresponding to all degrees of freedom of the unit cell, are equal.

{\bf Remark.} Heat transport in several scalar lattices with initial temperature distribution~\eq{T0 sin} is considered in papers~\cite{Krivtsov DAN 2015, Krivtsov 2019 ballistic, Kuzkin JPhys}. In the present paper, we derive a general solution valid for any lattice described by equations of motion~\eq{EM matr}.

The temperature matrix at time~$t$ is calculated by  formula~\eq{TT gen spher}. Substitution
of initial temperature profile~\eq{T0 sin} into~\eq{TT gen spher} after some transformations yields
\be{fast and slow sin}
\begin{array}{l}
 \DS \TT_F = \frac{1}{2}\(T_b + \Delta T\sin\frac{2\pi x}{L}\)\Matr{F}(t),
 \qquad
 \TT_S = \frac{T_b}{2} \EE + \frac{\Delta T}{2}\Matr{S}(t) \sin\frac{2\pi x}{L},
\\[4mm]
\DS
\Matr{F} = \int_{\kk}
\PP \tilde{\Matr{F}} \PP^{*\top}  {\rm d} \kk,
 \quad
\Matr{S} =  \int_{\kk}
\PP \tilde{\Matr{S}} \PP^{*\top}  {\rm d} \kk,
   \qquad
 \DS
 \tilde{F}_{ij} = \cos(2\omega_j t) \delta_{ij} ,
 \quad
 \tilde{S}_{ij} = \cos \frac{2\pi {v}_g^j t}{L} \delta_{ij}.
  \end{array}
\ee
Here $v_g^j = \Vect{v}_g^j \cdot \Vect{e}$.
Calculating trace in formula~\eq{fast and slow sin} we obtain simple expression for kinetic temperature:
\be{fast and slow sin tr}
\begin{array}{l}
 \DS T = T_F + T_S,
 %\\[4mm]
 \qquad
  \DS T_F = \frac{1}{2N}\(T_b+ \Delta T\sin\frac{2\pi x}{L}\)
  \sum_{j=1}^N \int_{\kk} \cos(2\omega_j t)  {\rm d} \kk,
  \\[4mm]
  \DS T_S = \frac{T_b}{N} + \frac{\Delta T}{2N} \sin\frac{2\pi x}{L} \sum_{j=1}^N\int_{\kk} \cos\frac{2\pi {v}_g^j t}{L}  {\rm d} \kk .
  \end{array}
\ee

Formula~\eq{fast and slow sin tr} shows that temperature profile remains sinusoidal at any moment in time. Therefore we compute amplitude of a sin as a function of time. In real experiment, the amplitude can be measured using the transient thermal grating technique~\cite{Johnson Temp Grat, Rogers Therm Grating}.
In one-dimensional case the amplitude is calculated as
\be{A def}
  \Matr{A} = \frac{2}{L} \int_{0}^{L} \TT(x,t) \sin\frac{2\pi x}{L} {\rm d}x,
  \qquad
  A = \frac{1}{N} \tr\Matr{A} =\frac{2}{L} \int_{0}^{L} T(x,t) \sin\frac{2\pi x}{L} {\rm d}x,
\ee
To increase the accuracy, in two-, and three-dimensional cases, results are additionally integrated in directions, orthogonal to~$x$.
Substituting expression for temperature~\eq{fast and slow sin}
into formula~\eq{A def}, yields
\be{A analyt}
  \Matr{A} = \Delta T \(\Matr{F}(t)  + \Matr{S}(t) \),
  \qquad
  A =  \frac{\Delta T}{2N} \sum_{j=1}^N \int_{\kk} \(\cos(2\omega_j t)  + \cos\frac{2\pi {v}_g^j t}{L}\) {\rm d} \kk.
\ee
Formula~\eq{A analyt} shows that time evolution of amplitude~$A$ depends on  direction,~$\Vect{e}$, of initial thermal perturbation. Therefore heat transport in two- and three-dimensional lattices is generally anisotropic~(see, for example, figure~\ref{graphene_sin}).

Formula~\eq{fast and slow sin} shows that $\TT_F$ and $\TT_S$ have {\it different time scales}, proportional to~$1/\omega_j$ and $L/v_g^j$ respectively. The first time scale is determined
by frequencies of vibrations of individual atoms. The second
time scale is determined by a time required for a wave to travel
distance~$L$. The ratio of these time scales is a large parameter. Therefore time scales of fast and slow thermal processes are well separated.

From formula~\eq{A analyt} and the stationary phase method~\cite{Fedoryuk} it follows that amplitude, $A$, of temperature profile in ballistic regime decays as~$1/t^\frac{d}{2}$. Note that solution of analogous problem using Fourier and hyperbolic~(Maxwell-Cattaneo-Vernotte~\cite{Chandrasekharaiah hyperbolic, Tzou hyperbolic}) heat transfer equations yields exponential decay of the amplitude.

Thus decay of amplitude of sinusoidal temperature profile
in a purely ballistic regime is described by formulas~\eq{A analyt}.
Presented results may serve for interpretation of experimental data obtained by transient thermal grating technique~\cite{Johnson Temp Grat, Rogers Therm Grating}. In particular, in a recent paper~\cite{Huberman graphite experiment} it is shown experimentally that decay of a sinusoidal profile in polycrystalline graphite at temperatures about~$100 K$ is nonmonotonic. Similar effect is predicted by our formula~\eq{A analyt}~(see e.g. figure~\ref{graphene_sin}).

% ---------------------------------------------------------------------------------------------
% ---------------------------------------------------------------------------------------------
% ---------------------------------------------------------------------------------------------
% ---------------------------------------------------------------------------------------------

\section{Example. Diatomic chain}\label{sect Diatomic}
In this section, ballistic heat transport in the  simplest one-dimensional polyatomic lattice is analyzed. We demonstrate that formulas~\eq{fast and slow app} describe time evolution of a temperature profile with high accuracy. Also we show that during heat transport temperatures, corresponding to two degrees of freedom of the unit cell, are generally different even if their initial values are equal.

\subsection{Equations of motion}
\begin{figure*}[htb]
\begin{center}
\includegraphics*[scale=0.35]{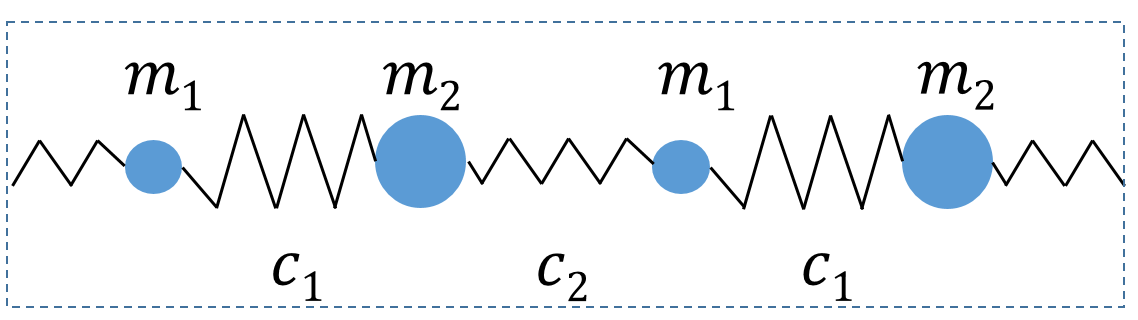}
\caption{Two unit cells of a diatomic chain with alternating masses and stiffnesses. Particles of different size form two sublattices.}
\label{chain_diff_mass_stiff}
\end{center}
\end{figure*}

We consider a diatomic chain with alternating masses~$m_1$, $m_2$ and stiffnesses~$c_1$,
$c_2$~(see fig.~\ref{chain_diff_mass_stiff}). The chain consists of two {\it sublattices}, one formed by particles with mass~$m_1$
and another formed by particles with mass~$m_2$.

We write equations of  motion of the chain in matrix form~\eq{EM matr}.
Unit cells, containing two particles each, are numbered by index~$j$. Position vector of the unit cell~$j$ has form
\be{}
\xx_j= x_j\Vect{e}, \qquad x_j= a \(j- \frac{n_c}{2}\),
\ee
where $a$ is a distance between unit cells; $\ve$ is a unit vector directed along the chain; $n_c$ is the total number of unit cells in the periodic cell. Each particle has one degree of freedom.
Displacements of particles, belonging to the unit cell~$j$, form a column
\be{}
  \u_j = \u(\xx_j) = \BM u_{1j} & u_{2j}\EM^{\top},
\ee
where $u_{1j},  u_{2j}$  are displacements of particles with masses~$m_1$ and $m_2$ respectively.
Then equations of motion have form
\be{B alp chain}
\begin{array}{l}
\MM\ddot{\u}_j = \CC_1 \u_{j+1} + \CC_0 \u_{j} + \CC_{-1} \u_{j-1},
\\[4mm]
 \DS \MM = \BM
    m_{1}   &  0 \\
    0       & m_{2}
    \EM,
\quad
\CC_0 =
\BM
-c_1-c_2   &  c_1 \\
c_1 &  -c_1-c_2
\EM,
\quad
\CC_1 = \CC_{-1}^{\top}= \BM
0   &  0 \\
c_2 &  0
\EM.
\end{array}
\ee

Initially particles have random velocities and zero displacements. In this section we consider isotropic initial temperature matrices, i.e.~$\TT_0(x_j)=T_0(x_j)\EE$. Then initial temperatures of the sublattices are equal~($T_{11}^0 = T_{22}^0 = T_0$). Corresponding
initial conditions for the particles have form:
\be{IC chain gen}
u_{1j} = u_{1j} = 0,
\quad
    \dot{u}_{1j} = \beta_{j}\sqrt{\frac{k_B}{m_1} T_0(x_j)},
    \quad
    \dot{u}_{2j} = \gamma_{j}\sqrt{\frac{k_B}{m_2} T_0(x_j)},
\ee
where $\beta_{j}$, $\gamma_{j}$ are uncorrelated random values with zero mean and unit variance,
i.e.~$\av{\beta_j} =\av{\gamma_j}= 0$, $\av{\beta_j^2} =\av{\gamma_j^2}= 1$, $\av{\beta_i\gamma_j}=0$ for all $i,j$.

\subsection{Dispersion relation and group velocities}
In this subsection, we calculate the dispersion relation and matrix~$\PP$ included in formulas~\eq{fast and slow app} for the temperature matrix.

We calculate  dynamical matrix,~$\Omb$, by formula~\eq{Omega}. Substituting expressions~\eq{B alp chain}
for matrixes~$\MM, \CC_{\alpha}$, $\alpha=0, \pm 1$ into formula~\eq{Omega}, we obtain:
\be{Omb chain1}
 \Omb =
 \BM
\frac{c_1+c_2}{m_1}   &  -\frac{c_1 +c_2 e^{-\ii p}}{\sqrt{m_1m_2}} \\
 -\frac{c_1 +c_2 e^{\ii p}}{\sqrt{m_1m_2}}    &  \frac{c_1+c_2}{m_2}
\EM,
\qquad
\kk = p \tilde{\Vect{b}},
\qquad
\tilde{\Vect{b}} = \frac{\ve}{a},
\ee
where $\kk$ is a wave vector; $p \in [0; 2\pi]$. Calculation of eigenvalues of matrix~$\Omb$,
 yields the dispersion relation:
%ïîëó÷àþòñÿ â ðåçóëüòàòå
%ðåøåíèÿ óðàâíåíèÿ:
%
%\be{disp chain2m}
% \omega^4 - \omega^2 (c_1 + c_2) \(\frac{1}{m_1} + \frac{1}{m_2}\)
% + 2 \frac{c_1c_2}{m_1m_2} \(1 - \cos p\) = 0.
%\ee
%
%Ðåøàÿ óðàâíåíèå~\eq{disp chain2m}, ïîëó÷èì äèñïåðñèîííîå ñîîòíîøåíèå~(ñîáñòâåííûå ÷èñëà ìàòðèöû $\Omb$):
%
\be{disp rel chain 2m}
\begin{array}{l}
\DS
\omega^2_{1,2}(p)
=
 \frac{\omega_{max}^2}{2}
  \(1 \pm \sqrt{1 - \frac{16 c_1c_2 \sin^2\frac{p}{2}}{ m_1m_2\omega_{max}^4}}\),
 \quad
  \omega_{max}^2 = \frac{(c_1+c_2)(m_1+ m_2)}{m_1m_2},
 \end{array}
\ee
where index~$1$ corresponds to plus sign.
Functions~$\omega_{1}(p), \omega_{2}(p)$ are referred to as optical and acoustic branches of the dispersion relation respectively.
Note that~$\omega_{1,2}/\omega_{max}$ equally depend on~$m_1/m_2$ and~$c_1/c_2$.

Group velocities are calculated by definition~\eq{def group vel}.
Projection of group velocities on direction of the chain for~$p \in (0; 2\pi)$ have form
\be{}
  v_g^{j} = a\frac{{\rm d} \omega_{j}}{{\rm d}p},
  \qquad
   v_g^{1} = \frac{c_1 c_2a \sin p}{m_1m_2\omega_1(\omega_1^2-\omega_2^2)},
  \qquad
  v_g^{2} = \frac{c_1 c_2a \sin p}{m_1m_2\omega_2(\omega_1^2-\omega_2^2)}.
\ee
Here~$\omega_{j}$ is a non-negative frequency in formula~\eq{disp rel chain 2m}. The maximum group velocity
is as follows
\be{max gr chain}
   v_* = \max_{p, j}|v_g^{j}| = a\sqrt{\frac{c_1 c_2}{(c_1+c_2)(m_1 + m_2)}}.
\ee

We calculate matrix~$\PP$ in equations~\eq{fast and slow app}. By the definition, columns of matrix~$\PP$ are equal to normalized eigenvectors of dynamical matrix~$\Omb$. Eigenvectors~$\Matr{d}_{1,2}$, corresponding to eigenvalues~$\omega_{1}^2, \omega_{2}^2$,
have form:
\be{PP chain}
 \Matr{d}_{1,2} = \BM 1-\frac{m_1}{m_2} \pm \sqrt{\(1-\frac{m_1}{m_2}\)^2 + 4|g|^2\frac{m_1}{m_2}} & -2g\sqrt{\frac{m_1}{m_2}}\EM^{\top},
    \qquad
 g = \frac{c_1 + c_2e^{\ii p}}{c_1 + c_2}.
\ee
Normalization of vectors~$\Matr{d}_{1,2}$ yields columns of matrix~$\PP$.

In the following sections, formulas~\eq{disp rel chain 2m}, \eq{PP chain} are employed for calculation of temperatures of sublattices~$T_{11}, T_{22}$.

\subsection{Thermal contact of cold and hot parts of the chain}\label{sect thermal contact chain}
In this section, we consider thermal contact of cold and hot parts of the chain and show that temperatures of
sublattices in this problem are different even though their initial values are equal.

Initial spatial distribution of
temperature matrix has form
\be{ICs Heaviside chain}
  \TT_0(x) = T_0(x)\EE, \qquad T_0(x) = T_b +  \Delta T H(x).
\ee
According to formula~\eq{ICs Heaviside chain} initial temperatures of sublattices are equal.
In our calculations~$\Delta T=T_b$. Since the chain is harmonic,
the value~$\Delta T/T_b$ does not change results qualitatively\footnote{Note that for anharmonic crystals~$\Delta T/T_b$ is an important
parameter of the problem, which can change results significantly.}.

Analytical solution of this problem is given by formulas~\eq{TT gen spher}, \eq{TfTs Heaviside}. Integrals in formulas~\eq{TT gen spher}, \eq{TfTs Heaviside} are evaluated numerically using Riemann sum approximation.
Interval of integration is divided into~$2 \cdot 10^4$ equal segments.

\begin{figure*}[htb]
\begin{center}
\includegraphics*[scale=0.45]{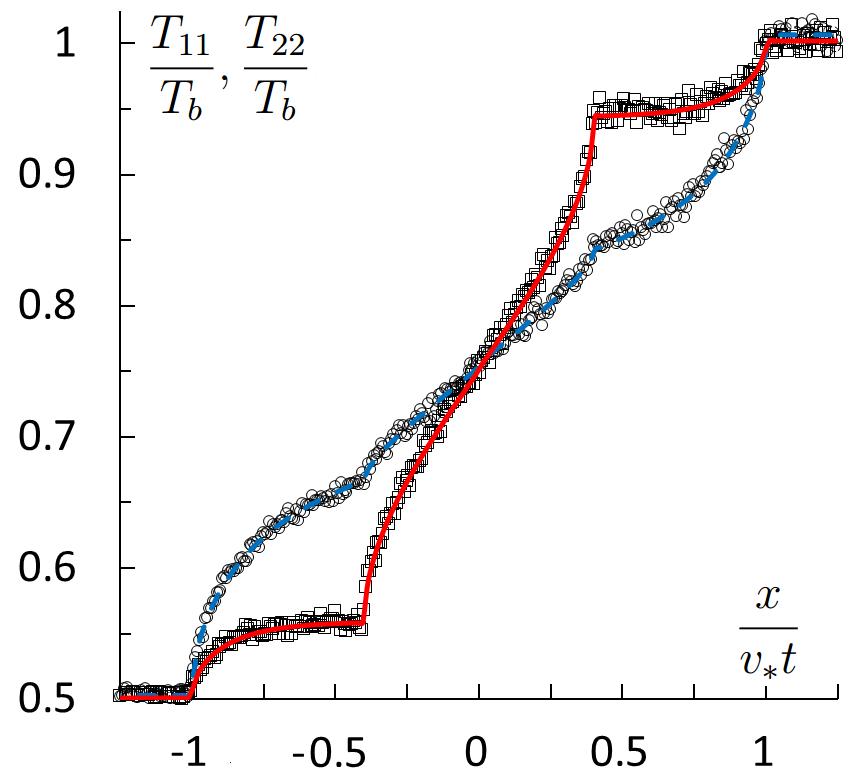}
\caption{Thermal contact of hot and cold parts of the diatomic chain~($m_2=2m_1, c_1=c_2$).
 Temperatures of sublattices~$T_{11}$~(solid red line and squares) and $T_{22}$ (dashed blue line and circles) at $t=500\tau_{min}$ are shown. Lines correspond to analytical solution~\eq{TfTs Heaviside}. Squares and triangles are results of numerical solution of equations of motion.}
\label{Heaviside_chain_m1m2}
\end{center}
\end{figure*}

To check formulas~\eq{TT gen spher}, \eq{TfTs Heaviside}, we compare them with results of numerical solution of equations of motion~\eq{B alp chain} with initial conditions~\eq{IC chain gen}, \eq{ICs Heaviside chain}. Numerical integration is carried out using
 symplectic  leap-frog integrator with time-step~$5 \cdot 10^{-3} \tau_{min}$.
 According to  formulas~\eq{TT gen spher}, \eq{TfTs Heaviside} temperature matrix at large times is self-similar, i.e. it depends on~$x/t$ only.
 Therefore it is sufficient to compare numerical and analytical results at a single moment in time.
 We compare results at~$t = 500 \tau_{min}$,
 where $\tau_{min} = 2\pi/\omega_{max}$, $\omega_{max}$ is defined by formula~\eq{disp rel chain 2m}.
 The chain consists of~$10^{4}$ unit cells under periodic boundary conditions.  During the simulation, kinetic temperatures of sublattices~$T_{11}, T_{22}$, at each unit cell, $j$, are calculated as
 \be{def T11 T22 chain}
   k_B T_{11}(x_j) = m \av{\dot{u}_{1j}^2}_r, \qquad  k_B T_{22}(x_j) = m \av{\dot{u}_{2j}^2}_r,
 \ee
where~$\av{...}_r$ stands for averaging over realizations of random initial conditions.
In the present example, number of realizations is equal to~$7 \cdot 10^4$.
Resulting temperatures of sublattices~$T_{11}, T_{22}$ at~$t=500 \tau_{min}$
for $m_2=2m_1, c_1=c_2$
are shown in figure~\ref{Heaviside_chain_m1m2}.
%
%$$
%\frac{T_{11}}{T_b}, \frac{T_{22}}{T_b}, \qquad \frac{x}{v_* t}
%$$

The figure shows that numerical results are accurately described by our approximate analytical formulas~\eq{TT gen spher}, \eq{TfTs Heaviside}.
 It is seen that temperatures of sublattices in the central part of the plot are {\it different} even though initially they are equal everywhere. This fact is predicted by formulas~\eq{TT gen spher},
\eq{TfTs Heaviside}.

{\bf Remark.} Difference of temperatures of light and heavy particles  has also been observed in {\it steady} problems considered in papers~\cite{Dhar altern mass, Jou diff mass}. Our results suggest that in unsteady problems light and heavy  particles also have generally different temperatures.

\subsection{Sinusoidal initial temperature profile}
In the present section, we consider decay of sinusoidal temperature profile~\eq{T0 sin} in a diatomic chain and show that, as in a previous example, temperatures of sublattices, $T_{11}, T_{22}$, at each unit cell are generally different. Decay of amplitude of sign is nonmonotonic.

Initial spatial distribution of temperature matrix has form
\be{IC sin chain}
    \TT_0(x) = T_0(x)\EE, \qquad T_0(x)= T_b + \Delta T \sin\frac{2\pi x}{L},
\ee
where $L$ is length of a periodic cell; in further calculations~$\Delta T=T_b/2$. Note that according to formula~\eq{IC sin chain} initial temperatures of sublattices are equal.
Analytical solution of this problem is given by formula~\eq{fast and slow sin}. The solution shows that spatial distribution remains sinusoidal at any
moment in time. Therefore we compute matrix~$\Matr{A}$ by formula~\eq{A def}. Elements~$A_{11}, A_{22}$ of this matrix correspond to amplitudes of temperatures~$T_{11}, T_{22}$.

Analytical expression for~$\Matr{A}$
is given by formula~\eq{A analyt}.
Integrals in formula~\eq{A analyt} are evaluated numerically using Riemann sum approximation.
Interval of integration is divided into~$10^3$ equal segments.
Below we compare predictions of this formula with results of numerical solution of equations of motion.

\begin{figure*}[htb]
\begin{center}
\includegraphics*[scale=0.44]{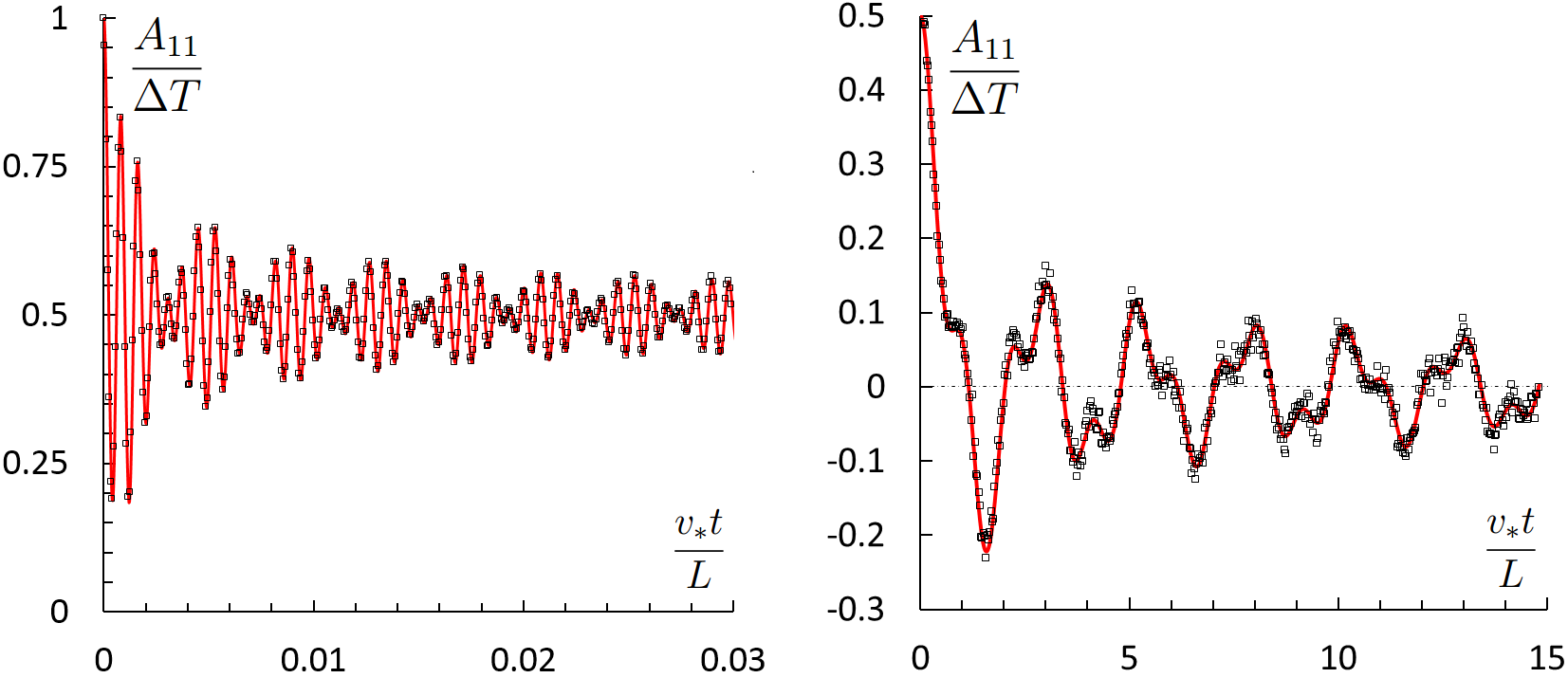}
\caption{Amplitude,~$A_{11}$, of sinusoidal temperature profile in a diatomic chain~($m_2=2m_1, c_1=c_2$) at short times~(left) and large times~(right). Analytical solution~\eq{A analyt}~(line) and numerical solution of equations of motion~(squares). Here~$v_*$ is the maximum group velocity~\eq{max gr chain}.}
\label{A11 chain}
\end{center}
\end{figure*}

\begin{figure*}[htb]
\begin{center}
\includegraphics*[scale=0.44]{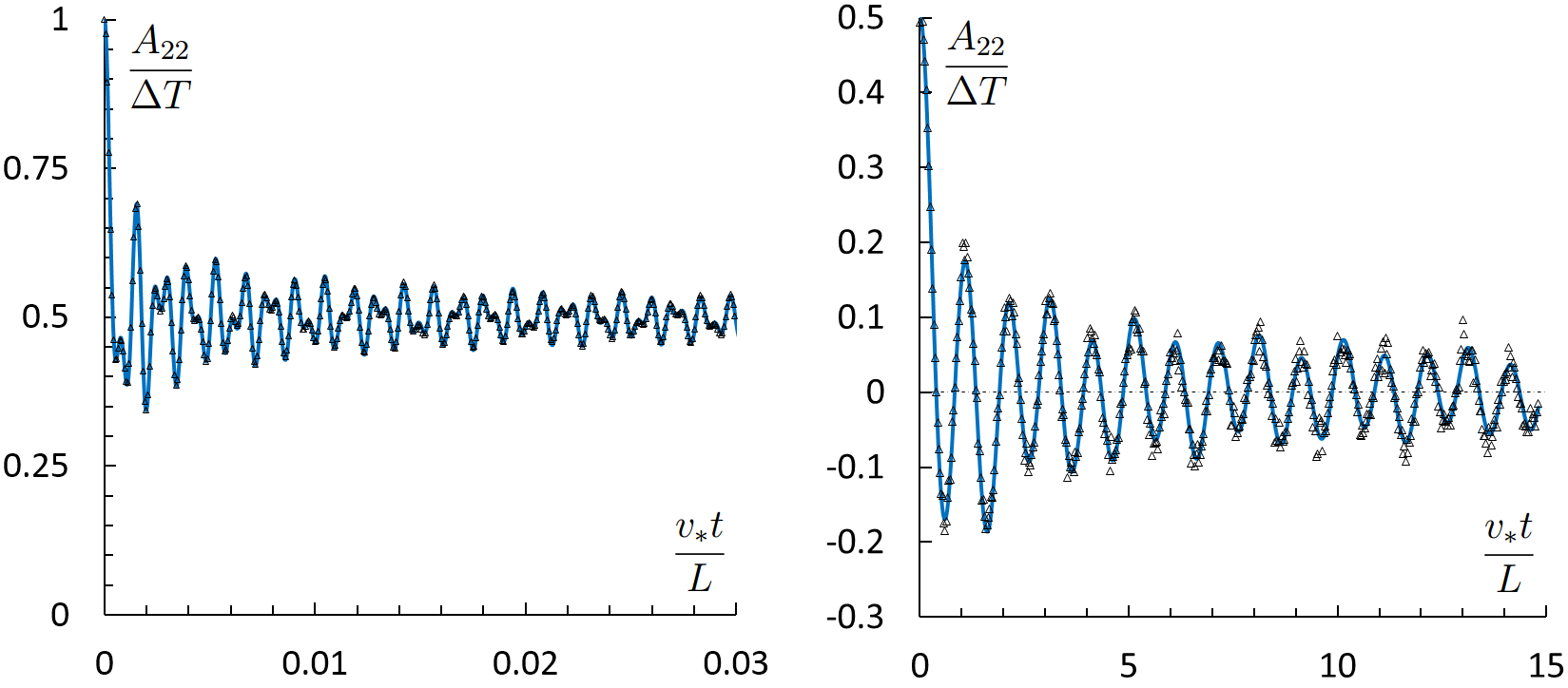}
\caption{Amplitude,~$A_{22}$, of sinusoidal temperature profile in a diatomic chain~($m_2=2m_1, c_1=c_2$) at short times~(left) and large times~(right). Analytical solution~\eq{A analyt}~(line) and numerical solution of equations of motion~(triangles). Here~$v_*$ is the maximum group velocity~\eq{max gr chain}.}
\label{A22 chain}
\end{center}
\end{figure*}

In computer simulations, the chain consists of~$10^{4}$ unit cells under periodic boundary conditions. Equations of motion~\eq{B alp chain} are solved numerically with initial conditions~\eq{IC chain gen}, \eq{IC sin chain}. During the simulation matrix~$\Matr{A}$ is calculated by
formula~\eq{A def},
where integral is replaced by summation with respect to all unit cells. Temperatures of unit cells are calculated
by formula~\eq{def T11 T22 chain}.
Resulting value of~$\Matr{A}$ is averaged over~$10^2$ realizations with random initial conditions. Note that number of realizations is less than in the previous example, because formula~\eq{A def} involves additional spatial
averaging, increasing the accuracy.
Amplitudes~$A_{11}, A_{22}$ of temperatures~$T_{11}, T_{22}$ for $m_2=2m_1, c_1=c_2$ are shown 
in figures~\ref{A11 chain}, \ref{A22 chain}.
%
%$$
%  \frac{A_{11}}{\Delta T}  \frac{A_{22}}{\Delta T} \qquad\frac{ v_*t}{L}
%$$
Figures show that numerical results are accurately described by our approximate analytical formula~\eq{A analyt}.

Thus, similarly to the previous example, temperatures of sublattices for $t>0$
are {\it different}~($T_{11} \neq T_{22}$), while their initial values are equal. We also note that decay of amplitude of sinusoidal temperature profile is {\it nonmonotonic}.

% ---------------------------------------------------------------------------------------------
% ---------------------------------------------------------------------------------------------
% ---------------------------------------------------------------------------------------------
% ---------------------------------------------------------------------------------------------

\section{Example. Graphene (out-of-plane motions)} \label{sect graphene}
In this section, we consider ballistic heat transport in
graphene lattice~(see fig.~\ref{graphene_lattice}). Only out-of-plane vibrations are considered. The model describes out-of-plane vibrations of a stretched graphene sheet~\cite{Balandin graphene, Berinskii, Hizhnyakov graphene}. In-plane vibrations can be considered separately, since in harmonic approximation in-plane and out-of-plane vibrations are decoupled. The main goal of this section is to show that approximate formulas~\eq{fast and slow app} describe behavior of temperature in two-dimensional lattices with high accuracy. Additionally we demonstrate individual contributions of acoustic and optical vibrations to thermal transport.

\subsection{Equations of motion}

In this subsection we represent equations of motion for the graphene lattice in a matrix form.

\begin{figure*}[htb]
\begin{center}
\includegraphics*[scale=0.36]{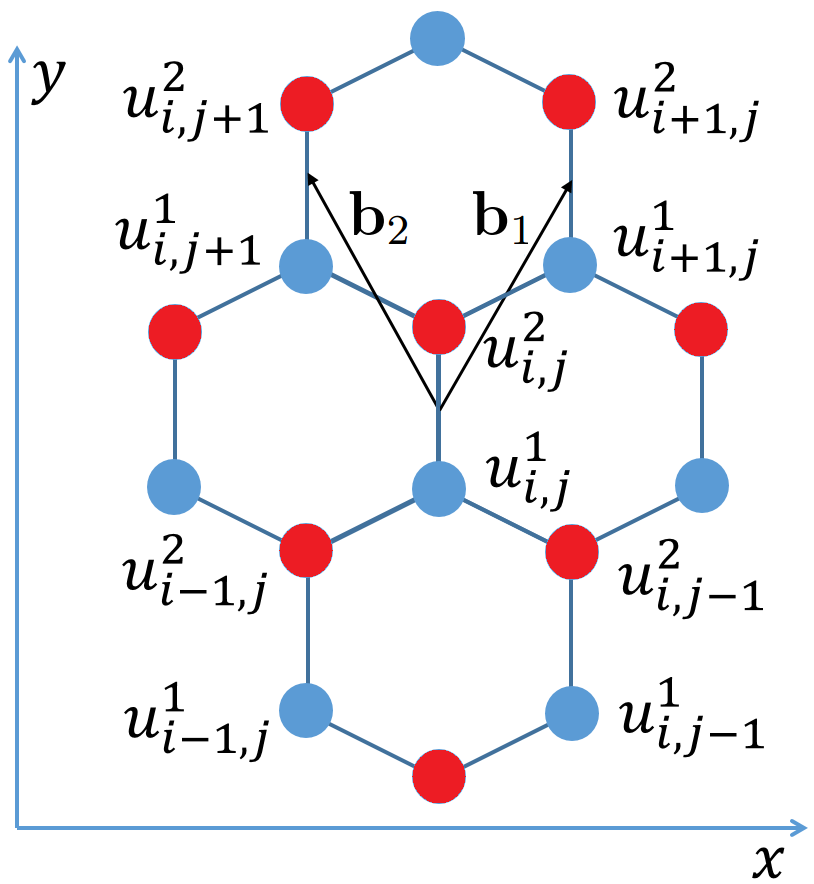}
\caption{Numbering of unit cells in graphene lattice. Here~$\Vect{b}_1$, $\Vect{b}_2$ are primitive vectors of the lattice. Particles move along the normal to lattice plane. $x$, $y$ axes correspond to zigzag and armchair directions respectively.}
\label{graphene_lattice}
\end{center}
\end{figure*}

The lattice is shown in figure~\ref{graphene_lattice}. Unit cells, containing two particles each,
are numbered by pair of indices~$i,j$.
%
% ----------------------------------------------------
% ----------------------------------------------------
%
Primitive vectors~$\Vect{b}_1, \Vect{b}_2$ for the lattice have form
\be{b1 b2 graphene}
   \Vect{b}_1 = \frac{\sqrt{3}a}{2} \( \Vect{i} + \sqrt{3}\Vect{j} \),
   \qquad
   \Vect{b}_2 = \frac{\sqrt{3}a}{2} \(\sqrt{3}\Vect{j} - \Vect{i}\),
\ee
where $\Vect{i}, \Vect{j}$ are Cartesian  unit vectors directed along $x$ and $y$ axes respectively; $a$ is equilibrium distance between the nearest particles. Vector~$\Vect{b}_1$ connects centers of cells~$i,j$ and $i+1,j$.
 Vector~$\Vect{b}_2$ connects centers of cells~$i,j$ and $i,j+1$. Position vector of cell~$i,j$ is represented
 in terms of the primitive vectors as
\be{}
  \xx_{i,j} = i \Vect{b}_1 + j  \Vect{b}_2.
\ee

Each particle has one degree of freedom~(displacement normal to lattice plane).
 Displacements of a unit cell~$i,j$ form a column:
\be{}
 \u_{i,j} = \u(\xx_{i,j}) = \BM u^1_{i,j} &  u^2_{i,j}\EM^{\top},
\ee
where~$u^1_{i,j}, u^2_{i,j}$ are displacements of two sublattices.

Consider equations of motion of unit cell~$i,j$. Each particle is connected with three nearest neighbors by linear springs~(solid lines in fig.~\ref{graphene_lattice}).
Equilibrium length of the spring is less than initial distance between particles, i.e. the graphene sheet is uniformly
stretched~\footnote{In the absence of stretching, out-of-plane vibrations are essentially nonlinear. Nonlinear effects in unstrained graphene are considered e.g. in paper~\cite{Dmitriev graphene}.}. Stiffness of the spring, determined by stretching force, is denoted by~$c$.
Then equations of motion have form
\be{Ba graphene}
\begin{array}{l}
\MM\ddot{\u}_{i,j}
=
\CC_{2} \u_{i,j+1} +
\CC_{1} \u_{i+1,j} 
+ \CC_{0} \u_{i,j}
+
\CC_{-1} \u_{i-1,j}
 +  \CC_{-2} \u_{i,j-1},
\\[4mm]
\DS
\MM = m\EE,
\quad
\CC_0 =
\BM
-3c &  c \\
c   &  -3c
\EM,
\quad
\CC_1 = \CC_2 = \BM
0  &  0 \\
c  &  0
\EM,
.
\end{array}
\ee
Here~$\CC_{-1} = \CC_{1}^{\top}$, $\CC_{-2} = \CC_{2}^{\top}$; $m$ is mass of a particle.

Initially particles have random velocities and zero displacements. In this section we consider isotropic initial temperature
matrices:
\be{}
\TT_0(x_{i,j}, y_{i,j}) =  T_0(x_{i,j}, y_{i,j})\EE,
\ee
where~$x_{i,j}, y_{i,j}$ are Cartesian coordinates of position vector~$\xx_{i,j}$. In this case initial
temperatures of the sublattices are equal~($T_{11}^0 = T_{22}^0 = T_0$). Corresponding initial conditions for the particles have form:
\be{IC graph gen}
u_{i,j}^1 = u_{i,j}^2 = 0,
\quad
    \dot{u}_{i,j}^1 = \beta_{i,j}\sqrt{\frac{k_B }{m}T_0(x_{i,j}, y_{i,j})},
    \quad
    \dot{u}_{i,j}^2 = \gamma_{i,j}\sqrt{\frac{k_B}{m} T_0(x_{i,j}, y_{i,j})},
\ee
where $\beta_{i,j}$, $\gamma_{i,j}$ are uncorrelated random values with zero mean and unit variance, i.e. $\av{\beta_{i,j}} =\av{\gamma_{i,j}}= 0$, $\av{\beta_{i,j}^2} =\av{\gamma_{i,j}^2}= 1$, $\av{\beta_{i,j}\gamma_{s,p}}=0$ for all~$i,j,s,p$.

Further we consider time evolution of kinetic temperature~$T = \frac{1}{2}\(T_{11} + T_{22}\)$.

\subsection{Dispersion relation and group velocities}
In this subsection, we calculate the dispersion relation, matrix~$\PP$~(see formula~\eq{PP def}), and group velocities.

We calculate dynamical matrix~$\Omb$ using formula~\eq{Omega}. Substituting expressions~\eq{Ba graphene} for matrixes~$\MM, \CC_{\alpha}$,
$\alpha=0;\pm1; \pm2$ into formula~\eq{Omega}, yields
\be{Omb graphene}
\begin{array}{l}
\DS
 \Omb = \omega_*^2
            \BM
            3                   &  -1 -e^{-\ii p_1}-e^{-\ii p_2} \\
            -1 -e^{\ii p_1}-e^{\ii p_2}   &  3
            \EM,
            \qquad
            p_1 = \kk\cdot \Vect{b}_1, \quad p_2 = \kk\cdot \Vect{b}_2,
\end{array}
\ee
where~$\kk$ is wave-vector; $\omega_*^2 = \frac{c}{m}$;
$p_1, p_2 \in [0; 2\pi]$ are dimensionless components of the wave vector.

Eigenvalues~$\omega_{1}^2, \omega_{2}^2$ of matrix~$\Omb$ determine dispersion relation for the lattice.
Solution of the eigenvalue problem yields:
\be{disp gr}
    \omega_{1,2}^2 = \omega_*^2
    \(3 \pm R(p_1,p_2)\),
    \qquad
    R(p_1,p_2) = \sqrt{3 + 2 \(\cos p_1 + \cos p_2 + \cos\(p_1-p_2\)\)},
\ee
where index~$1$ corresponds to plus sign.
Functions~$\omega_{1}(p_1,p_2)$, $\omega_{2}(p_1,p_2)$ are referred to as optical and acoustic dispersion surfaces
respectively.
Eigenvectors of matrix~$\Omb$ are columns of matrix~$\PP$:
\be{PP graphene}
 \PP =
    \frac{1}{\sqrt{|b|^2 + b^2}}
    \BM
    |b|   &  |b| \\
    -b    &   b
    \EM,
 \qquad
 b = 1 + e^{\ii p_1} + e^{\ii p_2}.
\ee

Group velocities~$\Vect{v}_{g}^{1}, \Vect{v}_{g}^{2}$ for $p_1, p_2 \in (0; 2\pi)$ are calculated by definition~\eq{def group vel} as
\be{cg graphene}
  \begin{array}{l}
    \DS  \Vect{v}_{g}^{j} = \frac{\partial \omega_{j}}{\partial \kk}
   =
  \frac{\partial \omega_{j}}{\partial p_1} \Vect{b}_1
  + \frac{\partial \omega_{j}}{\partial p_2} \Vect{b}_2,
  \\[4mm]
  \DS
  \frac{\partial \omega_{1}}{\partial p_1}
  =
  \frac{-\omega_*^2 \(\sin p_1 + \sin(p_1-p_2)\)}{2\omega_{1}R(p_1,p_2)},
  \qquad
  \frac{\partial \omega_{2}}{\partial p_1}
  =
  \frac{\omega_*^2 \(\sin p_1 + \sin(p_1-p_2)\)}{2\omega_{2}R(p_1,p_2)},
  \\[4mm]
  \DS
  \frac{\partial \omega_{1}}{\partial p_2}
  =
  \frac{-\omega_*^2 \(\sin p_1 - \sin(p_1-p_2)\)}{2\omega_{1}R(p_1,p_2)},
  \qquad
  \frac{\partial \omega_{2}}{\partial p_2}
  =
  \frac{\omega_*^2 \(\sin p_1 - \sin(p_1-p_2)\)}{2\omega_{2}R(p_1,p_2)}.
  \end{array}
\ee
Here~$\omega_{1} \geq 0, \omega_{2} \geq 0$; function~$R(p_1,p_2)$ is defined by formula~\eq{disp gr}; primitive vectors~$\Vect{b}_1,  \Vect{b}_2$ are given by formula~\eq{b1 b2 graphene}.

According to formulas~\eq{cg graphene}, the maximum absolute values of group velocities corresponding to acoustic and optical branches are as follows
\be{max vg fraphene}
 \max_{p_1,p_2} |\Vect{v}_{g}^{1}| \approx 0.448v_*, 
 \qquad 
 \max_{p_1,p_2} |\Vect{v}_{g}^{2}| \approx 0.897v_*, 
 \qquad 
 v_*=\omega_*a.
\ee

In the following sections, formulas~\eq{Omb graphene}, \eq{disp gr},  \eq{PP graphene}, \eq{cg graphene} are employed for description of ballistic heat transport.

\subsection{Circular initial temperature profile}
In this subsection we consider contributions of acoustic and optical vibrations to ballistic heat transport.

Initially the temperature has constant value, $T_1$, inside a circle 
of radius~$R$ and vanishes outside:
\be{circ distr}
\TT_0 = T_0(x,y)\EE,
\qquad
 T_0(x,y) =
 \begin{cases}
    T_1, & x^2 + y^2 \leq R^2, \\
    0, & x^2 + y^2 > R^2,
 \end{cases}
\ee
In  our calculations~$R=10a$. Analytical solution of this problem is calculated using formula~\eq{T tr}. Integrals in formula~\eq{T tr} are evaluated numerically using Riemann sum approximation. Integration area is divided into~$300\times 300$ equal square elements.

\begin{figure*}[htb]
\begin{center}
\includegraphics*[scale=0.27]{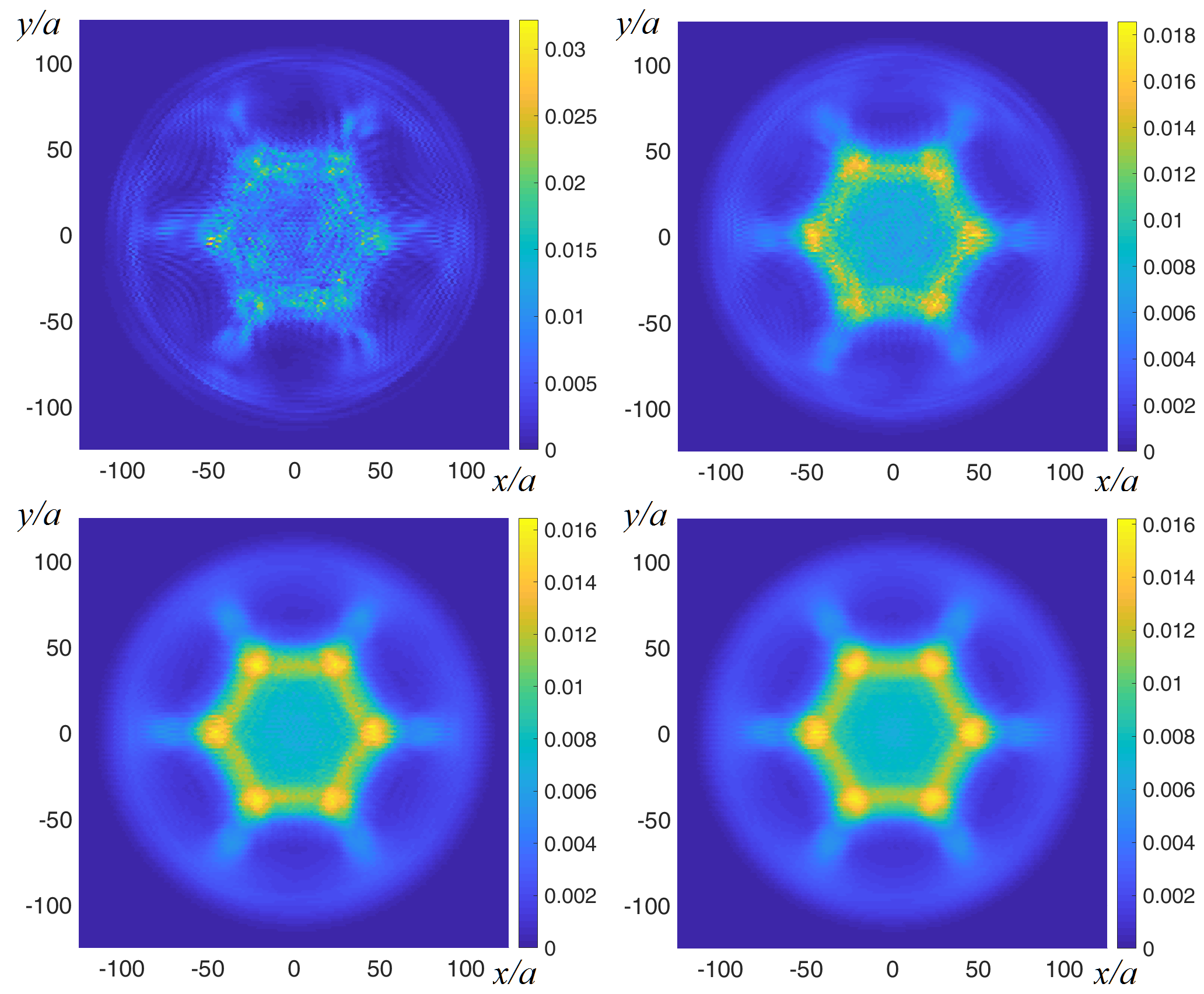}
\caption{Temperature field in graphene at~$t=20\tau_*$ corresponding to circular distribution of initial temperature~\eq{circ distr}. Results of numerical solution of lattice dynamics equations averaged over~$10, 10^2, 10^3, 10^4$ realizations are shown. Color bars show~$T/T_1$, where $T_1$ is constant in formula~\eq{circ distr}.}
\label{graphene_delta_all}
\end{center}
\end{figure*}

In computer simulations a square graphene sheet of length~$L=300a$ is considered. Equations of lattice dynamics~\eq{Ba graphene} with initial conditions~\eq{IC graph gen}, \eq{circ distr} are solved numerically using leap-frog integrator with
time-step~$5 \cdot 10^{-3}\tau_*$, $\tau_*=2\pi/\omega_*$.
 Kinetic temperatures of all unit cells~$T(x_{i,j}, y_{i,j})$ at~$t=20 \tau_*$  are calculated as
\be{}
 k_B T(x_{i,j}, y_{i,j}) = \frac{1}{2} m \av{\(\dot{u}^1_{i,j}\)^2 + \(\dot{u}^2_{i,j}\)^2}_r,
\ee
where averaging is carried out with respect to realizations of random initial conditions.
The moment in time is chosen such that fast relaxation process can be neglected.
Resulting temperature fields averaged over~$10, 10^2, 10^3, 10^4$
realizations are shown in figure~\ref{graphene_delta_all}.
With increasing number of realizations, results of numerical solution of equations of motion converges
to analytical solution given by  formula~\eq{T tr}. For~$10^4$ realizations plots of analytical and numerical solutions
are visually  indistinguishable.

Figure~\ref{graphene_delta_all} shows, in particular, that heat front is a circle as predicted by formula~\eq{front eq}. At the same time, the temperature field has a symmetry of the lattice, i.e. the heat transport is significantly anisotropic.

\begin{figure*}[htb]
\begin{center}
\includegraphics*[scale=0.5]{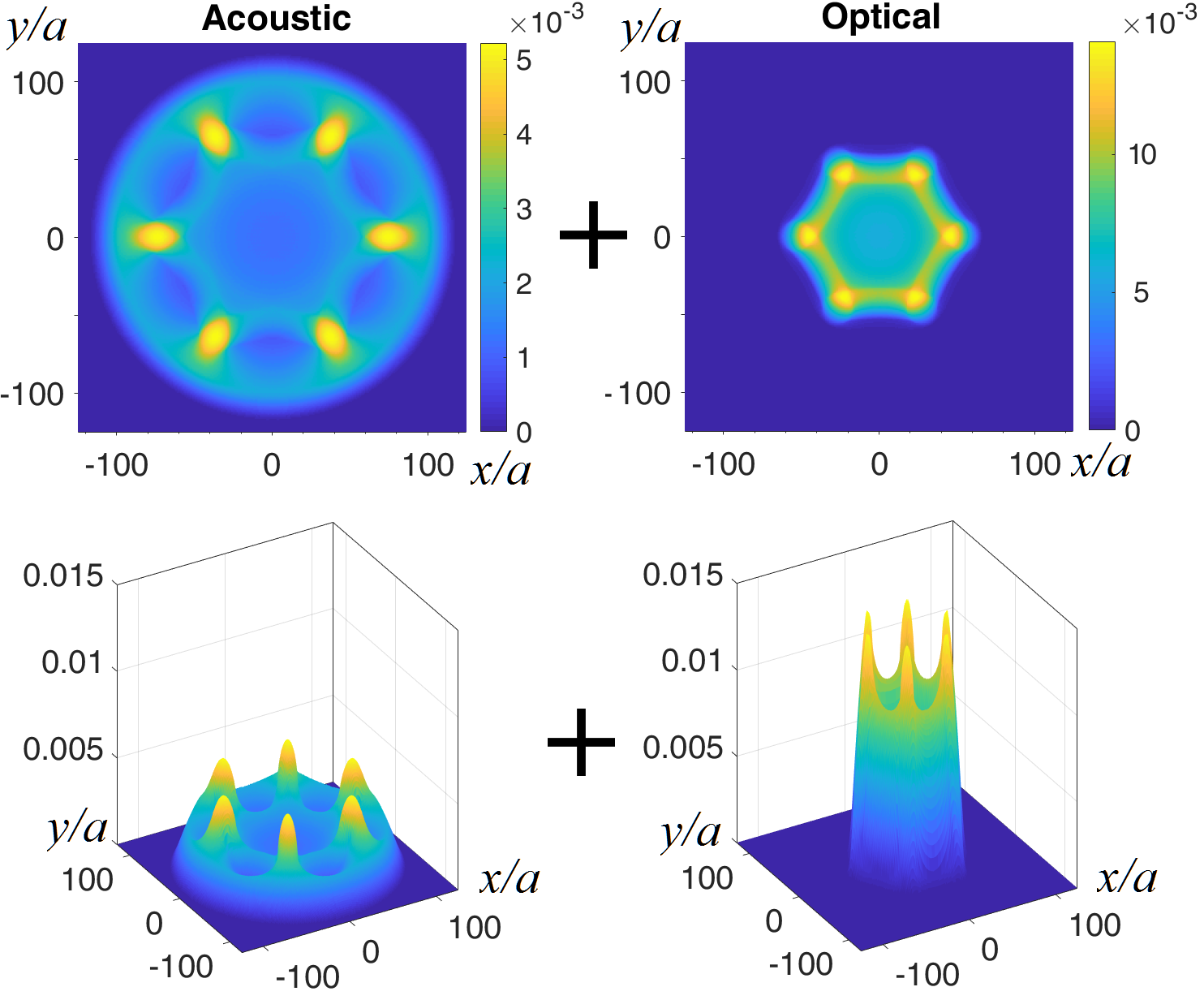}
\caption{Contribution of  acoustic~(left) and optical~(right) vibrations to temperature field in graphene at~$t=20\tau_*$ for circular distribution of initial temperature~\eq{circ distr} with~$R=10a$.  Color bars show~$T/T_1$, where $T_1$ is constant in formula~\eq{circ distr}. Plus signs mean that the resulting temperature field can be obtained by summation of acoustic and optical contributions.}
\label{graphene acoustic vs optical}
\end{center}
\end{figure*}

According to formula~\eq{T tr}, temperature field has contributions from acoustic and optical
branches of dispersion relation. The contributions are shown in figure~\ref{graphene acoustic vs optical}.
Acoustic waves have larger group velocities than optical waves. Therefore temperature front on the left plot from figure~\ref{graphene acoustic vs optical} propagates faster. 

Thus formula~\eq{T tr} describes temperature distribution in graphene lattice with high accuracy. In contrast to numerical results, formula~\eq{T tr}  allows to analyze individual contributions of acoustic and optical
vibrations to ballistic heat transport.

\subsection{Thermal contact of cold and hot half-planes} \label{sect thermal contact graphene}
In this subsection we consider thermal contact of two half-planes with initial temperatures~$T_b$ and $2T_b$ (see formula~\eq{ICs Heaviside}). Temperatures of sublattices are equal. Since thermal transport in graphene is anisotropic, we consider two problems
with temperature distribution in $x$ and $y$ directions:
\be{IC Heaviside graphene}
    \TT_0 = T_0(x,y)\EE,
    \qquad
    T_0(x,y) = T_b\(1  +  H(x)\)\quad \mathrm{or} \quad  T_0(x,y) = T_b\(1  + H(y)\).
\ee
Analytical solution of this problem is given by formula~\eq{TfTs Heaviside trace}.

To check formula~\eq{TfTs Heaviside trace}, we compare them with results of numerical solution of equations of motion. Formula~\eq{TfTs Heaviside trace} shows that at large times the solution is self-similar. Therefore it is sufficient to consider temperature field at a single moment in time. In our calculations it is equal to~$t= 20 \tau_{*}$.
Nearly square graphene sheet containing $301 \times 348$ unit cells is considered. Particles have
random initial velocities corresponding to initial temperature
distributions~\eq{IC Heaviside graphene}. Initial particle displacements are equal
to zero.
\begin{figure*}[htb]
\begin{center}
\includegraphics*[scale=0.4]{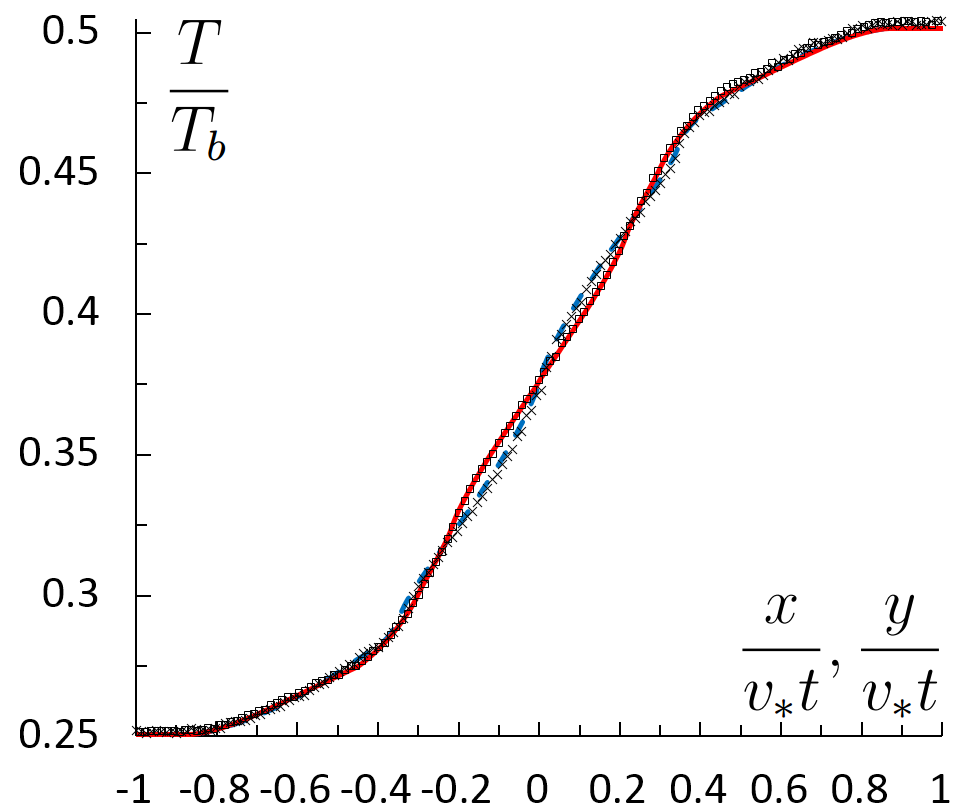}
\caption{Thermal contact of hot and cold parts of graphene.
 Solutions of two problems with temperature distributions in $x$~(solid red line) and $y$~(dashed blue line)
  directions at~$t=19.5\tau_{*}$ are shown. Analytical solution~\eq{TfTs Heaviside trace}~(lines)
  and results of numerical solution of equations of motion~(squares and crosses). Here~$v_* = \omega_*a$.}
\label{Heaviside_graphene}
\end{center}
\end{figure*}
Periodic boundary conditions in both directions are
used. During the simulation, kinetic energy of each unit cell is calculated. In order to calculate temperatures, results are averaged over $1.5\cdot10^3$ realizations with random initial conditions.
 Temperature distributions in $x$ and $y$ directions are shown in figure~\ref{Heaviside_graphene}.

The figure~\ref{Heaviside_graphene} shows that numerical results are accurately described by our approximate formulas~\eq{fast and slow app}. 
%$$
%    \frac{T}{T_b} \qquad \frac{x}{v_*t}, \frac{y}{v_*t}
%$$

\subsection{Sinusoidal initial temperature profile}
In the present section, we consider decay of spatially sinusoidal temperature profile~\eq{T0 sin}
in graphene. 
We investigate the influence of lattice anisotropy on heat transfer by 
comparing solutions of two problems with temperature changing zigzag~($x$) 
and armchair~($y$) directions:
\be{IC sin graphene}
\TT_0 = T_0(x,y)\EE,
    \qquad
    T_0(x,y) = T_b + \Delta T \sin\frac{2\pi x}{L}
    \quad
    \mathrm{or}
    \quad
    T_0(x,y) = T_b + \Delta T \sin\frac{2\pi y}{L},
\ee
where $L$ is length of a periodic cell. In our calculations~$\Delta T=T_b/2$.

Analytical solution of this problem is given by formula~\eq{fast and slow sin}. The solution shows
that spatial distribution remains sinusoidal at any
moment in time. Therefore we compute amplitude of temperature profile~(see formula~\eq{A def}).
Analytical expression for~$A$ is given by the second formula from~\eq{A analyt}.
Integrals in formula~\eq{A analyt} are evaluated numerically using Riemann sum approximation.
Interval of integration is divided into~$200 \times 200$ equal segments.
Below we compare predictions of this formula with results of numerical solution of equations of motion.

We check the accuracy of formula~\eq{A analyt} using numerical
solution of equations of motion. Particles have
random initial velocities corresponding to initial temperature
distributions~\eq{IC sin graphene}. Initial particle displacements are equal
to zero. Periodic boundary conditions in both directions are
used. The periodic cell contains $200 \times 232$ unit cells.
During simulation amplitude,~$A$, is calculated using two-dimensional version of formula~\eq{A def}:
\be{A def 2D}
  A = \frac{2}{L^2} \int_{0}^{L}\int_{0}^{L} T(x,y) \sin\frac{2\pi x}{L} {\rm d}x {\rm d}y \quad \mathrm{or} \quad A = \frac{2}{L^2} \int_{0}^{L}\int_{0}^{L} T(x,y) \sin\frac{2\pi y}{L} {\rm d}x.
\ee
Integral in formula~\eq{A def 2D}  is replaced by sum with respect to unit cells.

\begin{figure*}[htb]
\begin{center}
\includegraphics*[scale=0.44]{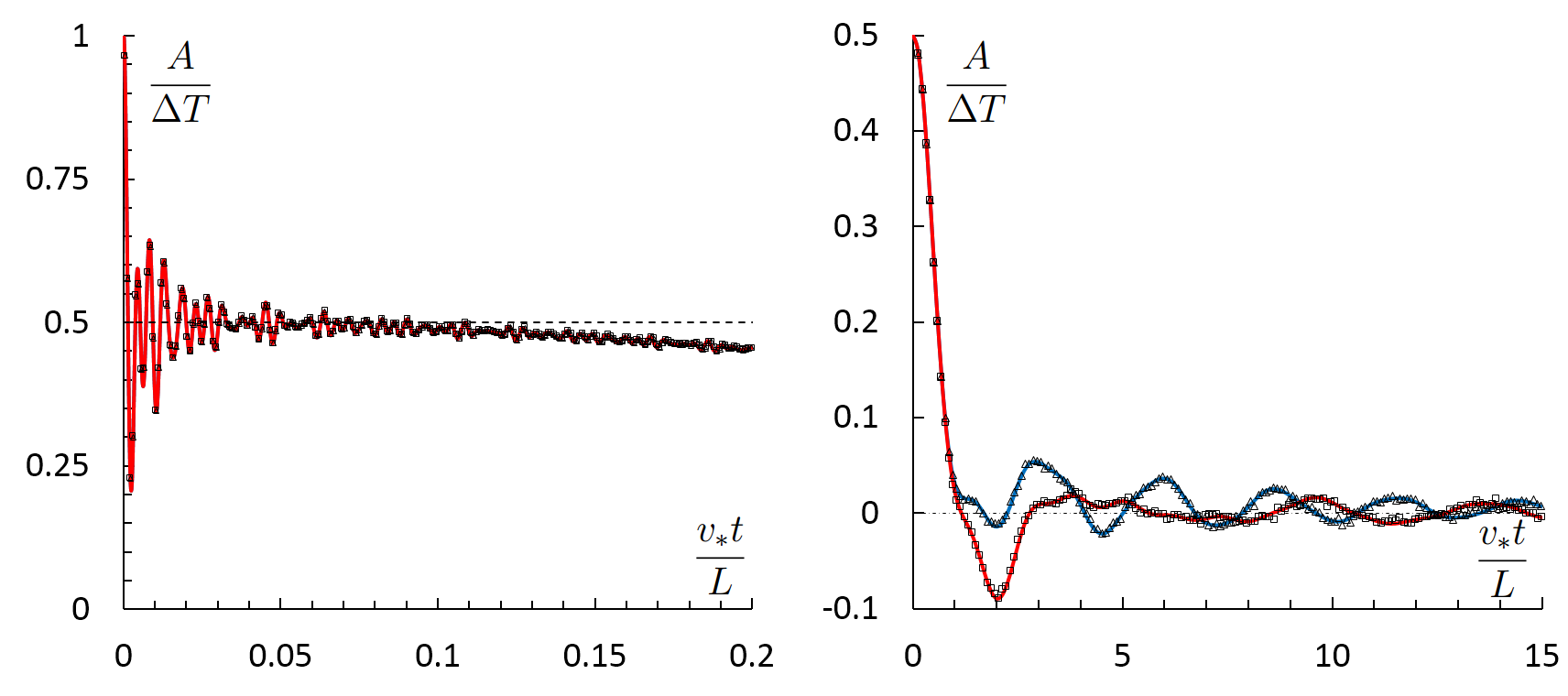}
\caption{Amplitude, $A$, of sinusoidal temperature profile in graphene at short times~(left) and large times~(right). Solutions of two problems with temperature distributions in zigzag~(red line) and armchair~(blue line) directions are shown. Analytical solution~\eq{A analyt}~(lines) and numerical solution of equations of motion~(squares and triangles). Here~$v_* = \omega_*a$.}
\label{graphene_sin}
\end{center}
\end{figure*}

Dependence of  dimensionless amplitude, $A/\Delta T$, on dimensionless time, $c_* t/L$, is
shown in figure~\ref{graphene_sin}. Every circle on the plot corresponds to average over realizations.
Figure~\ref{graphene_sin} shows that analytical solution~\eq{A analyt} practically coincides with results
of numerical solution of lattice dynamics equations. Results for~$x$ and $y$ directions practically coincide for~$t \leq L/v_*$, while for~$t > L/v_*$ they are significantly different.

Thus our analytical solution~\eq{A analyt} shows that decay of amplitude of the sinusoidal profiles in zigzag and armchair directions is {\it nonmonotonic}. Similar nonmonotonic behavior has been recently observed experimentally in polycrystalline graphite~\cite{Huberman graphite experiment} at temperatures about~$T_b \sim 100 K$ and length scales~$L \sim 1 \mu m$. The nonmonotonic behavior shows that at these time and length scales the heat transport in graphite is ballistic.

%$$
%\frac{A}{\Delta T}, \frac{A}{\Delta T} \qquad\frac{ v_*t}{L}
%$$

\section{Conclusions}
We have shown that time evolution of initial temperature profile in infinite harmonic crystals is described by formula~\eq{fast and slow app} with high accuracy. 

It was shown that at short times temperatures, corresponding to degrees of freedom of the unit cell, oscillate and tend to generally different equilibrium values. The oscillations are caused by redistribution of energy between kinetic and potential forms and redistribution of energy between degrees of freedom of the unit cell. In infinite crystals\footnote{In the case of finite crystals, the phenomenon of thermal echo introduced in paper~\cite{Murachev} is observed.}, the oscillations practically vanish at times of order of~$100$ periods of atomic vibrations.

Evolution of temperature profile at large times is caused by ballistic heat transport. At large times temperature field is represented as a superposition of waves having a shape of initial temperature distribution and traveling with group velocities depending on the wave vector. Formula~\eq{fast and slow app} has the same property as equations of lattice dynamics, it is  invariant with respect to change~$t$ by $-t$.

It is noteworthy that temperatures, corresponding to degrees of freedom of the unit cell, at large times are generally neither equal to each other nor equal to their equilibrium values. Therefore thermal state of unit cells reached by thermal waves is {\it strongly nonequilibrium}~(see e.g. figures~\ref{Heaviside_chain_m1m2}, \ref{A11 chain}, \ref{A22 chain}). It was shown that these analytical findings are in a good agreement with results of numerical solution of equations of motion.

Though formula~\eq{fast and slow app} allows to calculate the temperature profile, it does not yield an equation describing ballistic heat transport. So far the closed equation has been derived only for several particular lattices~\cite{Babenkov2018, Krivtsov DAN 2015, Krivtsov 2019 ballistic}. Derivation of the general equation valid for all harmonic crystals would be an interesting extension of the present work.

Presented theory may serve for proper statement and interpretation of experiments on unsteady ballistic heat transport in crystals. Analytical solution of the problem with sinusoidal profile of initial temperature can be used for interpretation of results obtained by the transient thermal grating technique~\cite{Johnson Temp Grat, Rogers Therm Grating, Huberman graphite experiment}. In particular, the solution predicts nonmonotonic decay of the temperature profile, which was recently observed experimentally in graphite~\cite{Huberman graphite experiment}.

\section{Acknowledgements}
The author is deeply grateful to his supervisor A.M. Krivtsov for the general statement of the problem and continuous stimulating discussions. The author greatly appreciate comments by S.V. Gavrilov, M.A. Guzev, D.A. Indeitsev, and M.L. Kachanov.

The work was financially supported by the Russian Science Foundation
under grant No. 17-71-10213.

%\section{To do}
%\begin{itemize}
%\item Check derivation of fundamental solution. Is multiplier $1/(4N(2\pi)^d)$ correct?
%\item Check formula~\eq{max vg fraphene}.
%\item applications: interpretation of experiments~(e.g. TG), benchmarking MD, benchmarking BTE
%\end{itemize}

\appendix
\section{Approximate formula for the temperature matrix}

In this section, approximate formula~\eq{fast and slow app} for the temperature matrix is derived.
The derivation is based on the assumption that the main contribution to integrals~$\TT_F$, $\TT_S$
comes from the points such that~$\kk_1 \approx \kk_2$.

The expression for~$\TT_F$ in formula~\eq{fast and slow app}  is derived as follows. We introduce  new variables in formula~\eq{fast slow ex} for~$\TT_F$:
\be{p1 p2}
    \pp_1=\kk_1, \qquad \pp_2=\kk_1-\kk_2.
\ee
Jacobian of this transformation is equal to unity. Using periodicity of the integrand in formula~\eq{fast slow ex}, it can be shown that integration is carried out in the same domain as in formula~\eq{int k}. We assume that the main contribution to the integral~\eq{fast and slow app} comes form the points~$\pp_2 \approx 0$~($\kk_1\approx \kk_2$).
Then integrand is expanded into series with respect to~$\pp_2$. In particular, the following approximate
formulas are used:
\be{}
   \PP(\pp_1-\pp_2) \approx  \PP(\pp_1), 
   \qquad  
   \omega_i(\pp_1) \pm \omega_j(\pp_1-\pp_2) 
   \approx 
   \omega_i(\pp_1) \pm \omega_j(\pp_1),
\ee
where~$i \neq j$.
Then formula~\eq{fast slow ex} for $\TT_F$ reads
\be{TF prel}
\begin{array}{l}
\DS
\TT_F \approx \int_{\pp_1}\PP \int_{\pp_2}\sum_{\yy}
\tilde{\TT}_F(\yy, \pp_1)e^{\ii \pp_2 \cdot(\xx-\yy)} {\rm d} \pp_2  \PP^{*\top}{\rm d} \pp_1,  \qquad \PP = \PP(\pp_1),
\\[4mm]
 \DS
\{\tilde{\TT}_F\}_{ij} = \frac{1}{2}\{\PP^{*\top} \TT_0(\yy) \PP\}_{ij} \[\cos((\omega_i(\pp_1) + \omega_j(\pp_1))t)  +
%\right. \\[4mm]
%\left.+
(1-\delta_{ij})\cos((\omega_i(\pp_1) - \omega_j(\pp_1))t)\].
\end{array}
\ee
According to the definition of the discrete Fourier transform the following identity is satisfied:
\be{}
    \int_{\pp_2}\sum_{\yy}
    \tilde{\TT}_F(\yy, \pp_1)e^{\ii \pp_2 \cdot(\xx-\yy)} {\rm d} \pp_2
    = \tilde{\TT}_F(\xx, \pp_1).
\ee
Substituting this identity into formula~\eq{TF prel} yields the expression for~$\TT_F$ in formula~\eq{fast and slow app}.

To derive approximate expression for~$\TT_S$ in formula~\eq{fast and slow app} we introduce new variables~\eq{p1 p2} in formula~\eq{fast slow ex}. The integrand in formula~\eq{fast slow ex} is expanded into series with respect to~$\pp_2$. In particular, the following approximate
formulas are used:
\be{}
\PP(\pp_1-\pp_2) \approx  \PP(\pp_1),
\qquad
\omega_j(\pp_1) - \omega_j(\pp_1-\pp_2) \approx \pp_2 \cdot \Vect{v}_g^j(\pp_1),
\ee
where $\Vect{v}_g^j$ is the group velocity defined by
formula~\eq{fast and slow app}. 
Then formula~\eq{fast slow ex} for $\TT_S$ reads
\be{Ts temp}
\begin{array}{l}
\DS \TT_S \approx \int_{\pp_1}\int_{\pp_2}\sum_{\yy}
\PP \tilde{\TT}_S(\yy)\PP^{*\top} e^{\ii \pp_2 \cdot(\xx-\yy)}  {\rm d} \pp_2{\rm d} \pp_1, \qquad \PP=\PP(\pp_1),
\\[4mm]
\DS \{\tilde{\TT}_S\}_{ij} \approx   \frac{1}{2}\{\PP^{*\top} \TT_0(\yy) \PP\}_{ij} \delta_{ij}\cos(\pp_2 \cdot \Vect{v}_g^j(\pp_1) t).
\end{array}
\ee
Using identity~$2\cos(\pp_2 \cdot \Vect{v}_g^j t) = e^{\ii\pp_2 \cdot \Vect{v}_g^j t} + e^{-\ii\pp_2 \cdot \Vect{v}_g^j t}$ and
properties of the discrete Fourier transform, we show that 
\be{temp1}
     \int_{\pp_2}\sum_{\yy}\{\PP^{*\top} \TT_0(\yy) \PP\}_{jj} \cos(\pp_2 \cdot \Vect{v}_g^j t) e^{\ii \pp_2 \cdot(\xx-\yy)} {\rm d} \pp_2
     = \frac{1}{2}\{\PP^{*\top} \(\TT_0(\xx + \Vect{v}_g^j t) +  \TT_0(\xx - \Vect{v}_g^j t)\)\PP\}_{jj}.
\ee
Substituting formula~\eq{temp1} into formula~\eq{Ts temp}, yields the expression for~$\TT_S$ in formula~\eq{fast and slow app}.

More rigorous derivation of formula~\eq{fast and slow app} is beyond the scope of the present paper. In the present paper, we show that formula~\eq{fast and slow app} has high accuracy by comparison of analytical solutions of several problems with results of numerical solution of lattice dynamics equations~(see sections~\ref{sect Diatomic}, \ref{sect graphene}).

% ------------------------------------------------------------------------------% -----------------------------------------------------------------------------------------------------
% -----------------------------------------------------------------------------------------------------
% -----------------------------------------------------------------------------------------------------
% -----------------------------------------------------------------------------------------------------
% -----------------------------------------------------------------------------------------------------


\begin{thebibliography}{1}
%\bibitem{Allen1969} K.R. Allen, J. Ford, \paptit{Energy transport for a
%   three-dimensional harmonic crystal,} Phys. Rev., {\bf 187}, 1132 (1969).

\bibitem{Allenbook} M.P. Allen, D.J. Tildesley, {\it Computer Simulation of Liquids.} (Clarendon Press, Oxford, 1987), p. 385.

\bibitem{Anufriev nanowires} R. Anufriev, S. Gluchko, S. Volz, M. Nomura \paptit{Quasi-ballistic heat conduction due to lévy phonon flights in silicon nanowires}, ACS Nano, 2018, DOI: 10.1021/acsnano.8b07597

\bibitem{Babenkov2016}  M.B. Babenkov,  A.M. Krivtsov,  D.V. Tsvetkov.  \paptit{Energy oscillations in 1D harmonic crystal on elastic foundation.}  Phys. Mesomech. 2016, {\bf 19}, 1, pp. 60-67.

\bibitem{Babenkov2018} M.B. Babenkov,  A.M. Krivtsov,  D.V. Tsvetkov. \paptit{Heat propagation in the one-dimensional harmonic crystal on an elastic foundation.} Phys. Mesomech, 2019 [in press]

\bibitem{Balandin graphene} A.A. Balandin,  \paptit{Thermal properties of graphene and nanostructured carbon materials.}, Nat. Mat. {\bf 10} (2011).

 \bibitem{Bonetto2004} F. Bonetto, J.L. Lebowitz, J. Lukkarinen,
    \paptit{Fourier's law for a harmonic crystal with self-consistent
    stochastic reservoirs.} J. Stat. Phys.,  {\bf 116}, 783 (2004).

\bibitem{Dmitriev graphene} E. Barani, I.P. Lobzenko, E.A. Korznikova, E.G. Soboleva, S.V. Dmitriev, K. Zhou, A.M. Marjaneh \paptit{Transverse discrete breathers in unstrained graphene} Eur. Phys. J. B, Vol. 90, Is. 3, 1 (2017)

\bibitem{BGK} P.L. Bhatnagar, E.P. Gross, M. Krook \paptit{A model for collision processes in gases. I. Small amplitude processes in charged and neutral one-component systems}, Physical Review. 94 (3): 511–525, (1954)

%\bibitem{Benettin slow relax} G. Benettin, G. Lo Vecchio, A. Tenenbaum \paptit{Stochastic transition in two-dimensional Lennard-Jones systems} Phys. Rev. A, 22, 1709 (1980)

\bibitem{Berinskii} I.E. Berinskii, A.M. Krivtsov, Linear oscillations of suspended Graphene. In: Altenbach H., Mikhasev G. (eds) Shell and Membrane Theories in Mechanics and Biology. Advanced Structured Materials, vol 45. Springer.

\bibitem{Bonetto 2000} F. Bonetto, J. L. Lebowitz, and L. Rey-Bellet, Fourier's law: a challenge to theorists, in A. Fokas, A. Grigoryan, T. Kibble, and B. Zegarlinski (eds.), Mathematical Physics 2000, (Imperial College Press, London, 2000), pp. 128–150.

%\bibitem{Boldrighini1983} C. Boldrighini,  A. Pellegrinotti, L. Triolo \paptit{Convergence to Stationary States for Infinite Harmonic Systems}, J. Stat. Phys., {\bf 30}, 1, 123--155 (1983).

\bibitem{Cahill 2003 review} D.G. Cahill, W.K. Ford, K.E. Goodson, G.D. Mahan,
    A. Majumdar, H.J. Maris, R. Merlin, S.R. Phillpot, \paptit{Nanoscale
    thermal transport.} J. Appl. Phys. {\bf 93}, 793 (2003).

%\bibitem{Cases two temp review} J. Casas-Vazquez,  D. Jou, \paptit{Temperature in non-equilibrium states: a review of open problems and current proposals}, Rep. Prog. Phys. 66 (2003) 1937–2023

\bibitem{Gang Chen}  G. Chen, \paptit{Ballistic-diffusive heat conduction equations.} Phys. Rev. Lett., Vol. 85, 2001, pp. 2297-2300

%\bibitem{Casher Lebowitz} A. Casher, J. L. Lebowitz, \paptit{Heat Flow in Regular and Disordered Harmonic Chains}, J. Math. Phys. 12, 1701 (1971).

%\bibitem{Chang2016} A.Y. Chang , Y.-J. Cho , K.-C. Chen, C.-W. Chen , A. Kinaci, B.T. Diroll, M.J. Wagner , M.K. Y. Chan , H.-W. Lin, R.D. Schaller, \paptit{Slow Organic-to-Inorganic Sub-Lattice Thermalization in Methylammonium Lead Halide Perovskites Observed by Ultrafast Photoluminescence} Adv. Energy Mater. {\bf 6}, 1600422 (2016).

\bibitem{Chang nanotubes} C.W. Chang, D. Okawa, H. Garcia, A. Majumdar, A. Zettl,  \paptit{Breakdown of Fourier's law in nanotube thermal conductors.} Phys. Rev. Lett. 101, 075903 (2008)

\bibitem{Chang review exp} C.W. Chang, in: Thermal transport in low dimensions,     Lecture Notes in Physics, Vol. 921, 2016, pp. 305-338.

\bibitem{Chandrasekharaiah hyperbolic} D.S. Chandrasekharaiah, \paptit{Hyperbolic thermoelasticity: a review of recent literature.} Appl.  Mech. Rev 39, 355–376 (1986)

%\bibitem{Chien diff mass} C.-C. Chien, S. Kouachi, K.A. Velizhanin, Y. Dubi,  M. Zwolak, \paptit{Thermal transport in dimerized harmonic lattices: Exact solution, crossover behavior, and extended reservoirs}, Phys. Rev. E 95, 012137 (2017).

\bibitem{Dhar 2008} A. Dhar, \paptit{Heat transport in low-dimensional systems.} Adv. Phys., pp. 457-537,  2008.

\bibitem{DharSaito2016} A. Dhar, K. Saito, in: Thermal transport in low
    dimensions, Lecture Notes in Physics, Vol. 921, pp. 305-338 (2016).

\bibitem{Dove} M.T. Dove, {\it Introduction to lattice dynamics}. (Cambridge University Press, London, 1993).

%\bibitem{Dobrushin 1986} R. L. Dobrushin, A. Pellegrinotti, Yu.M. Suhov, L. Triolo  \paptit{One-Dimensional Harmonic Lattice Caricature
%of Hydrodynamics}, J. Stat. Physics, {\bf 43}, 3/4 (1986).

%\bibitem{Dudnikova_2003} T.V. Dudnikova, A.I. Komech, H. Spohn, \paptit{On the convergence to statistical equilibrium
%for harmonic crystals.}  J. Math. Phys. {\bf 44}, 2596 (2003).

%\bibitem{Dudnikova 2005} T. V. Dudnikova, A. I. Komech, \paptit{On the Convergence to a Statistical Equilibrium in the Crystal Coupled to a Scalar Field,} Russian J. Math. Phys. 12 (3), 301–325 (2005).

\bibitem{Fedoryuk}	 M.V. Fedoryuk \paptit{The stationary phase method and pseudodifferential operators.} Russian Mathematical Surveys, 6(1), 65-115, (1971).

\bibitem{Gavrilov 2018 - 1}  S.N. Gavrilov, A.M. Krivtsov, D.V. Tsvetkov \paptit{Heat transfer in a one-dimensional harmonic crystal in a viscous environment subjected to an external heat supply.} Cont. Mech. Thermodyn., 2018, DOI: 10.1007/s00161-018-0681-3

\bibitem{Guzev2018} M. A. Guzev, \paptit{The exact formula for the temperature of a one-dimensional crystal.}  Dal'nevost. Mat. Zh., {\bf 18}, 39 (2018)


\bibitem{Gendelman 2010 nonstat} O.V. Gendelman, A.V. Savin,
    \paptit{Nonstationary heat conduction in one-dimensional chains with
    conserved momentum.} Phys. Rev. E, {\bf 81}, 020103, (2010).

\bibitem{Harris} L. Harris, J. Lukkarinen, S. Teufel, F. Theil,
\paptit{Energy transport by acoustic modes of harmonic lattices.} SIAM J.
    Math. Anal., {\bf 40}(4) 1392 (2008).


\bibitem{Hizhnyakov graphene} V. Hizhnyakov, M. Klopov, A. Shelkan \paptit{Transverse intrinsic localized modes in monoatomic chain and in
graphene.} Phys. Let. A, Vol. 380, Is. 9–10, pp. 1075-1081 (2016).

%\bibitem{HEMMEN} J.L. van Hemmen \paptit{A generalized equipartition theorem}, Phys. Lett., Vol. 79A, No. 1 (1980)

%\bibitem{Hemmer} P.C. Hemmer, {\it Dynamic and stochastic types of motion in the linear chain.} (Norges tekniske hoiskole, 1959).

%\bibitem{Hoover Holian} B.L. Holian, W.G. Hoover, B. Moran, G.K. Straub. \paptit{Shock-wave structure via nonequilibrium molecular dynamics and Navier-Stokes continuum mechanics.}
% Phys. Rev. A, {\bf22}, 2798 (1980).

%\bibitem{Holian Mareschal} B.L. Holian, M. Mareschal. \paptit{Heat-flow equation motivated by the ideal-gas shock wave} Phys. Rev. E, {\bf 82}, 026707 (2010).

%\bibitem{Hoover_stat_phys} W.G. Hoover, {\it Computational statistical mechanics}, (Elsevier, N.Y., 1991). p. 330.

%\bibitem{Hoovers} W.G. Hoover, C.G. Hoover, K.P. Travis. \paptit{Shock-wave compression and Joule-Thomson expansion.} Phys. Rev. Lett., {\bf112}, 144504 (2014).

%\bibitem{Huerta1969} M.A. Huerta, H.S. Robertson \paptit{Entropy, Information Theory, and the Approach to Equilibrium of Coupled Harmonic Oscillator Systems}, J. Stat. Phys., {\bf 1}, 3, 393-414 (1969).

%\bibitem{Huerta Robertson 1971} M.A. Huerta, H.S. Robertson, J.C. Nearing, \paptit{Exact Equilibration of Harmonically Bound Oscillator Chains}, J. Math. Phys. 12, 2305 (1971).

\bibitem{Hoover thermostat} W.G. Hoover, C.G. Hoover, \paptit{Hamiltonian thermostats fail to promote heat flow.} Commun. Nonlinear Sci. Numer. Simulat. 18 3365, 2013.

\bibitem{Hoover_stat_phys} W.G. Hoover, {\it Computational statistical
    mechanics}, (Elsevier, N.Y., 1991). p. 330.

\bibitem{Hsiao ballistic} T.K. Hsiao, H.K. Chang, S.-C. Liou, M.-W. Chu, S.-C. Lee, C.-W. Chang, \paptit{Observation of room-temperature ballistic thermal conduction persisting over 8.3 $\mu$m SiGe nanowires.} Nat. Nanotech., {\bf
8(7)}, 534 (2013).

\bibitem{Hua Minnich 2014 BTE} C. Hua, A.J. Minnich \paptit{Transport regimes in quasiballistic heat conduction.} Phys. Rev. B 89, 094302 (2014).
    
\bibitem{Huberman graphite experiment}  S. Huberman, R.A. Duncan, K.Chen, B. Song, V. Chiloyan, Z. Ding, A.A. Maznev, G. Chen, K.A. Nelson, \paptit{Observation of second sound in graphite at temperatures above 100 K}, arXiv:1901.09160 [cond-mat.mes-hall], (2019)


\bibitem{Indeitsev2017} D.A. Indeitsev, E.V. Osipova, \paptit{A two-temperature model of optical excitation of acousticwaves in conductors.} Dokl. Phys. 62(3), 136–140 (2017)

\bibitem{Johnson Temp Grat} J.A. Johnson, A.A. Maznev, J. Cuffe, J.K. Eliason, A.J. Minnich, T.Kehoe, C.M. Sotomayor Torres, G. Chen, K.A. Nelson \paptit{Direct measurement of room-temperature nondiffusive thermal transport over micron distances in a silicon membrane.} Phys. Rev. Lett., 110, 025901 (2013).

%\bibitem{Petrov Fortov} N.A. Inogamov, Yu.V. Petrov, V.V. Zhakhovsky, V.A. Khokhlov, B.J. Demaske, S.I. Ashitkov, K.V. Khishchenko, K.P. Migdal, M.B. Agranat, S.I. Anisimov, V.E. Fortov, I.I. Oleynik.
%\paptit{Two-temperature thermodynamic and kinetic properties of transition metals irradiated by femtosecond lasers} AIP Conf. Proc. {\bf 1464}, 593 (2012)

\bibitem{Dhar altern mass} V. Kannan, A. Dhar, J.L. Lebowitz   \paptit{Nonequilibrium stationary state of a harmonic crystal with alternating masses.} Phys. Rev. E, {\bf 85}, 041118 (2012).

\bibitem{Jou diff mass}  A. Kato, D. Jou, \paptit{Breaking of equipartition in one-dimensional heat-conducting systems.} Phys. Rev. E, 64, 052201, (2001).

%\bibitem{Dmitriev2} L.Z. Khadeeva, S.V. Dmitriev, Yu.S. Kivshar \paptit{Discrete breathers in deformed graphene}, Jetp Lett. 94: 539 (2011).

%\bibitem{Huerta Robertson 1969} H.S. Robertson, M.A. Huerta, \paptit{APPROACH TO EQUILIBRIUM OF COUPLED, HARMONICALLY BOUND OSCILLATOR SYSTEMS}, Phys. Rev. Lett. 23, 825 (1969).

%\bibitem{Huerta Robertson 1969} M.A. Huerta,H.S. Robertson, \paptit{Entropy, Information Theory,  and the Approach to Equilibrium of Coupled Harmonic Oscillator Systems} J. Stat. Phys., Vol. 1, No. 3, 1969

%\bibitem{Klein Prigogine} G. Klein and I. Prigogine, \paptit{Sur la mecanique statistique des phenomenes irreversibles III}, Physica 19, 1053 (1953).

\bibitem{Klemens 1951} P.G. Klemens \paptit{The thermal conductivity of dielectric solids at low temperatures.} Proc. R. Soc. Lond. A, 208(1092), 108-133, 1951.

%\bibitem{Klemens 1958} P.G. Klemens, \paptit{Thermal Conductivity and Lattice Vibrational Modes}. Solid State Physics, 1–98, 1958.

\bibitem{Krivtsov 2014 DAN} A.M. Krivtsov \paptit{Energy oscillations in a one-dimensional crystal.}, Dokl. Phys., Vol. 59, No. 9, pp. 427--430, 2014.

\bibitem{Krivtsov DAN 2015} A.M. Krivtsov, \paptit{Heat transfer in infinite harmonic one dimensional crystals.} Dokl. Phys. {\bf 60(9)}, 407   (2015).

\bibitem{Krivtsov 2019 ballistic} A.M. Krivtsov, The ballistic heat equation for a one-dimensional harmonic crystal, in: {\it Dynamical processes in generalized continua and structures}, Springer Nature,  2019.


\bibitem{Koh BTE} Y.K. Koh, D.G. Cahill, B. Sun, \paptit{Nonlocal theory for heat transport at high frequencies.} Physical Review B, 90(20), 2014.

\bibitem{KosevichSavin} Y.A. Kosevich, A.V. Savin, \paptit{Confining
    interparticle potential makes both heat transport and energy diffusion
    anomalous in one-dimensional phononic systems.} Phys. Lett. A, {\bf 380},
    3480 (2016).

%\bibitem{Krivtsov 2003} A.M. Krivtsov \paptit{From nonlinear oscillations to equation of state in simple discrete systems}, Chaos, Solitons and Fractals {\bf 17}, 79–87 (2003)

%\bibitem{Kosevich} A.M. Kosevich, {\it The crystal lattice: phonons, solitons, dislocations, superlattices.} (John Wiley \& Sons, 2006).

\bibitem{Kubo 1973} R. Kubo, The Boltzmann equation in solid state physics. In: Cohen E.G.D., Thirring W. (eds) The Boltzmann Equation. Acta Physica Austriaca (Supplementum X Proceedings of the International Symposium ``100 Years Boltzmann Equation'' in Vienna 4th–8th September 1972), vol 10/1973. Springer, Vienna.

\bibitem{Kuzkin_DAN} V.A. Kuzkin, A.M. Krivtsov, \paptit{High-frequency thermal processes in harmonic crystals.} Dokl. Phys., {\bf 62(2)}, 85 (2017).

\bibitem{Kuzkin_FTT} V.A. Kuzkin, A.M. Krivtsov, \paptit{An analytical description of transient thermal processes in harmonic crystals.} Phys. Solid State, {\bf 59(5)}, 1051 (2017).

\bibitem{Kuzkin JPhys} V.A. Kuzkin, A.M. Krivtsov,  \paptit{Fast and slow thermal processes in harmonic scalar lattices.} J. Phys.: Condens. Matter, {\bf 29}, 505401 (2017).

\bibitem{Kuzkin_2018_arxive} V.A. Kuzkin, \paptit{Approach to thermal equilibrium in harmonic crystals with polyatomic lattice.} Cont. Mat. Thermodyn., 2019 [submitted].

%\bibitem{Lanford Lebowitz} O.E. Lanford, J.L.  Lebowitz,  Time evolution and ergodic properties of harmonic systems. In: Lecture Notes in Physics, Vol. 38, pp. 144--177. Berlin-Heidelberg-New York : Springer 1975

%\bibitem{Linn Robertson 1984} S.L. Linn, H.S. Robertson \paptit{Thermal energy transport in harmonic systems}, J. Phys. Chem. Sol., Vol. 45, 2, 1984, pp. 133-140.

%\bibitem{laser review} D. der Linde, K. Sokolowski-Tinten, J. Bialkowski, \paptit{Laser–solid interaction in the femtosecond time regime}, App. Surf. Sci., Vol. 109–110, 1997, pp. 1-10.

\bibitem{Lepri 2003} S. Lepri, R. Livi, A. Politi, \paptit{Thermal conduction in classical low-dimensional lattices.} Phys. Rep. {\bf 377}, 1 (2003).

\bibitem{Politi 2008} S. Lepri, C.  Mejia-Monasterio, A. Politi \paptit{A stochastic model of anomalous heat transport: analytical solution of the steady state.} J. Phys. A {\bf 42}, 2, 025001 (2008).

\bibitem{Krivtsov_LeZakharov} A.A. Le-Zakharov, A.M. Krivtsov,
    \paptit{Molecular dynamics investigation of heat conduction in crystals
    with defects.} Dokl. Phys., {\bf 53}, 261 (2008).

%\bibitem{Politi2010} S. Lepri, C. Mejia-Monasterio, A. Politi, J. Phys. A: Math., Theor. {\bf 43}, 065002 (2010).

%\bibitem{Marcelli slow relaxation} G.  Marcelli,  A. Tenenbaum \paptit{Quantumlike short-time behavior of a classical crystal}, Phys. Rev. E 68, 041112 (2003).

\bibitem{Mielke} A. Mielke, \paptit{Macroscopic behavior of microscopic
    oscillations in harmonic lattices via Wigner-Husimi transforms.} Arch.
    Ration. Mech. Anal., {\bf 181}, 401 (2006).

\bibitem{Minnich BTE} A.J. Minnich, G. Chen, S. Mansoor, B.S. Yilbas,  \paptit{Quasiballistic heat transfer studied using the frequency-dependent Boltzmann transport equation.} Phys. Rev. B, 84(23), p. 235207, 2011.

\bibitem{Mishuris} G.S. Mishuris, A.B.  Movchan, L.I. Slepyan, \paptit{Localised knife waves in a structured interface.} J. Mech. Phys. Solids, 57, 1958 (2009).

\bibitem{Murachev} A.S. Murachev, A.M. Krivtsov, D.V. Tsvetkov, \paptit{Thermal echo in a finite one-dimensional harmonic crystal.} J. Phys.: Cond. Mat., 31(9), 095702, (2019).

%\bibitem{Nakazawa1970} H. Nakazawa, \paptit{On the lattice thermal conduction.}     Progr. Phys., {\bf 45}, 231 (1970).

%\bibitem{Evans 2016} C.F. Petersen, D.J. Evans, S.R. Williams.
% \paptit{Dissipation in monotonic and non-monotonic relaxation to equilibrium.} J. Chem. Phys. {\bf144}, 074107 (2016).

\bibitem{Mahan nonlocal BTE} G.D. Mahan, F. Claro, \paptit{Nonlocal theory of thermal conductivity.} Phys. Rev. B 38, 1963, (1988).

%\bibitem{Simon book} S.H. Simon, {\it The Oxford solid state basics}. (OUP Oxford, 2013).

%\bibitem{Shredinger} E. Schrodinger,  Annalen der Physik, {\bf 44}, 916, (1914).

%\bibitem{Slepyan_equip}  L.I. Slepyan \paptit{On the
%energy partition in oscillations and waves}, Proc. R. Soc. A 471, 20140838 (2015).

\bibitem{Nika Balandin graphene} D.L. Nika, A.A. Balandin, \paptit{Two-dimensional phonon transport in graphene.} J. Phys.: Condens. Matter, 24, 233203 (2012).

\bibitem{Pierls 1929} R. Peierls, \paptit{Zur kinetischen theorie der warmeleitung in kristallen.} Ann. Phys. 3, 1055, (1929).

\bibitem{PiazzaLepri} F. Piazza, S. Lepri, \paptit{Heat wave propagation in a
    nonlinear chain.} Phys. Rev. B, {\bf 79}, 094306 (2009).

\bibitem{Poletkin2012} K.V. Poletkin, G.G. Gurzadyan, J. Shang, V. Kulish,
\paptit{Ultrafast heat transfer on nanoscale in thin gold films.} App. Phys. B, {\bf107}, 137 (2012).

 \bibitem{Pumarol ballistic}  M.E. Pumarol, M.C. Rosamond, P. Tovee, M.C. Petty, D.A. Zeze, V. Falko, O.V. Kolosov, \paptit{Direct nanoscale imaging of ballistic and diffusive thermal transport in graphene nanostructures.} Nano Lett., {\bf 12 (6)}, 2906 (2012).

%\bibitem{Prigogine normal modes} I. Prigogine, F. Henin \paptit{On the General Theory of the Approach to Equilibrium. I. Interacting Normal Modes}, J. Math. Phys. 1, 349 (1960)

\bibitem{Lebowitz1967} Z. Rieder, J.L. Lebowitz, E. Lieb, \paptit{Properties of a harmonic crystal in a stationary nonequilibrium state.} J. Math. Phys. {\bf 8}, 1073 (1967).

\bibitem{Romano Grossmann} G. Romano, J.C. Grossman, \paptit{Heat conduction in nanostructured materials predicted by phonon bulk mean free path distribution.} J. Heat Transfer, Vol. 137, 071302-1, (2015).

\bibitem{Rogers Therm Grating} J.A. Rogers, A.A. Maznev, M.J. Banet, K.A. Nelson, \paptit{Optical generation and characterization of acousticwaves in thin films: fundamentals and applications.} Annu. Rev. Mater. Sci., 30,  117–157 (2000).

\bibitem{Sokolov} A. A. Sokolov, A. M. Krivtsov, W. H. Muller \paptit{Localized heat perturbation in harmonic 1D crystals: Solutions for an equation of anomalous heat conduction.} Phys. Mesomech., {\bf 20}, 3, 305--310 (2017).

\bibitem{Sinha Goodson 2005 review BTE} S. Sinha, K.E. Goodson \paptit{Review: multiscale thermal modeling in nanoelectronics}, Int. J. Mult. Comp. Eng., 3(1), 107-133 (2005)

\bibitem{Spohn Lebowitz 1977} H. Spohn, J.L. Lebowitz, \paptit{Stationary non-equilibrium states of infinite harmonic systems.} Commun. math. Phys. {\bf 54}, 97 (1977).

\bibitem{Tsai nonstat} D.H. Tsai, R.A. MacDonald, \paptit{Molecular-dynamical
    study of second sound in a solid excited by a strong heat pulse.} Phys.
    Rev. B, {\bf 14(10)}, 4714 (1976).


\bibitem{Tzou hyperbolic} D.Y. Tzou, Macro- to microscale heat transfer: the lagging behavior, 566 p. Wiley (2015).

%\bibitem{Kuzkin Tsaplin} Tsaplin V.A., Kuzkin V.A. \paptit{Temperature oscillations in harmonic triangular lattice with random initial velocities} Lett. Mat., 8(1), 2018 pp. 16-20.

%\bibitem{Titulaer}  U.M. Titulaer, Physica 70, 257, 276, 456, (1973).

%\bibitem{Uribe shockwaves} F.J. Uribe, R.M. Velasco, L.S. Garcia-Colin,  \paptit{Two kinetic temperature description for shock waves.} Phys. Rev. E 58 3209–22, 1998.

\bibitem{Xiong2013 diff stiff}  D. Xiong, Y. Zhang, H. Zhao, \paptit{Heat transport enhanced by optical phonons in one-dimensional anharmonic lattices with alternating bonds.} Phys. Rev. E, 88, 052128 (2013).

%\bibitem{Ziman phonon} J.M. Ziman, {\it Electrons and Phonons. The theory of
%    transport phenomena in solids.} (Oxford University Press, New York, 1960),
%    p. 554.


%\bibitem{Zhilin} P.A. Zhilin, Vectors and second-rank tensors in three-dimensional space. St. Petersburg State Technical University. 1992,  86 p. (In Russian).

\bibitem{Xu Hu 2010 balst-diff BTE} M. Xu, H.Hu \paptit{A ballistic-diffusive heat conduction model extracted from Boltzmann transport equation.} Proc. Roy. Soc. A, Vol. 467, Is. 2131 (2010).


\bibitem{Xu grap exp 2014} X. Xu, L.F. Pereira, Y. Wang, ,J. Wu, K. Zhang, X. Zhao, S. Bae, C.T. Bui, R. Xie, J.T. Thong, B.H. Hong, K.P. Loh, D. Donadio, B. Li, B. Ozyilmaz, \paptit{Length-dependent thermal conductivity in suspended single-layer graphene.} Nat. Commun. 5, 3689 (2014).
\end{thebibliography}
\end{document}